\documentclass{aa} 

\usepackage[utf8]{inputenc}
\usepackage[varg]{txfonts}
\usepackage{graphicx}
\usepackage[font=small,labelfont=bf]{caption}
\usepackage{subfigure}
\usepackage{booktabs}
\usepackage{amsmath,amssymb}
\usepackage{gensymb}
\usepackage{wasysym}
\usepackage{scalerel}
\usepackage{natbib,twoopt}
\usepackage[flushleft]{threeparttable}
\usepackage{tabularx}
\usepackage{color}
\usepackage[breaklinks=true]{hyperref} 
\usepackage{adjustbox}
\usepackage{float}

\hypersetup{
      colorlinks=True,
      allcolors=blue
      }
\bibpunct{(}{)}{;}{a}{}{,} 
\makeatletter
\newcommandtwoopt{\citeads}[3][][]{\href{http://adsabs.harvard.edu/abs/#3}%
{\def\hyper@linkstart##1##2{}%
\let\hyper@linkend\@empty\citealp[#1][#2]{#3}}}
\newcommandtwoopt{\citepads}[3][][]{\href{http://adsabs.harvard.edu/abs/#3}%
{\def\hyper@linkstart##1##2{}%
\let\hyper@linkend\@empty\citep[#1][#2]{#3}}}
\newcommandtwoopt{\citetads}[3][][]{\href{http://adsabs.harvard.edu/abs/#3}%
{\def\hyper@linkstart##1##2{}%
\let\hyper@linkend\@empty\citet[#1][#2]{#3}}}
\newcommandtwoopt{\citeyearads}[3][][]%
{\href{http://adsabs.harvard.edu/abs/#3}
{\def\hyper@linkstart##1##2{}%
\let\hyper@linkend\@empty\citeyear[#1][#2]{#3}}}
\makeatother

\newcommand{\changeurlcolor}[1]{\hypersetup{urlcolor=#1}}

\def\kbol{{$k_{\rm bol, X}$}}
\def\ledd{{$\lambda_\mathrm{Edd}$}}
\def\lbol{{$\log{L_\mathrm{bol}}$}}

\def\ltsima{$\; \buildrel < \over \sim \;$}
\def\simlt{\lower.5ex\hbox{\ltsima}}

\begin{document}

  \title{X-ray spectroscopic survey of highly-accreting AGN}

  \titlerunning{X-ray spectroscopic survey of highly-accreting AGN}
  
  \subtitle{}

  \author{M.\ Laurenti\inst{1,2} \and E.\ Piconcelli\inst{2} \and L.\ Zappacosta\inst{2} \and F.\ Tombesi\inst{1,2,4,5} \and C.\ Vignali\inst{6,7} \and S.\ Bianchi\inst{8} \and P.\ Marziani\inst{9} \and F.\ Vagnetti\inst{1,3} \and A.\ Bongiorno\inst{2} \and M.\ Bischetti\inst{10} \and A.\ del Olmo\inst{11} \and G.\ Lanzuisi\inst{7} \and A.\ Luminari\inst{2,3} \and R.\ Middei\inst{2,12} \and M.\ Perri\inst{2,12} \and C.\ Ricci\inst{13,14} \and G.\ Vietri\inst{2,15}}
  
  \institute{Dipartimento di Fisica, Università di Roma ``Tor Vergata'', Via della Ricerca Scientifica 1, I-00133 Roma, Italy \\ \email{\changeurlcolor{black}\href{mailto:marco.laurenti@roma2.infn.it}{marco.laurenti@roma2.infn.it}} \and INAF - Osservatorio Astronomico di Roma, Via Frascati 33, I-00040 Monte Porzio Catone, Italy \and INAF - Istituto di Astrofisica e Planetologia Spaziali, Via del Fosso del Caveliere 100, I-00133 Roma, Italy \and Department of Astronomy, University of Maryland, College Park, MD 20742, USA \and NASA/Goddard Space Flight Center, Greenbelt, MD 20771, USA \and Dipartimento di Fisica e Astronomia ``Augusto Righi'', Università degli Studi di Bologna, Via Gobetti 93/2, I-40129 Bologna, Italy \and INAF – Osservatorio di Astrofisica e Scienza dello Spazio di Bologna, Via Gobetti 93/3, I-40129 Bologna, Italy \and Dipartimento di Matematica e Fisica, Università degli Studi Roma Tre, Via della Vasca Navale 84, I-00146 Roma, Italy \and INAF – Osservatorio Astronomico di Padova, Vicolo Osservatorio 5, I-35122 Padova, Italy \and INAF - Osservatorio Astronomico di Trieste, via G. B. Tiepolo 11, I–34143 Trieste, Italy \and Instituto de Astrof\'isica de Andaluc\'ia, IAA-CSIC, Glorieta de la Astronom\'ia s/n, 18008 Granada, Spain \and Space Science Data Center, SSDC, ASI, via del Politecnico snc, 00133 Roma, Italy \and N\'ucleo de Astronom\'ia de la Facultad de Ingenier\'ia, Universidad Diego Portales, Av. Ej\'ercito Libertador 441, Santiago 22, Chile \and Kavli Institute for Astronomy and Astrophysics, Peking University, Beijing 100871, People's Republic of China \and INAF - Istituto di Astrofisica Spaziale e Fisica Cosmica Milano, Via A.\ Corti 12, I-20133 Milano, Italy}

  \date{}

  \abstract
  {Improving our understanding of the nuclear properties of high-Eddington ratio ($\lambda_\mathrm{Edd}$) active galactic nuclei (AGN) is necessary since at this regime the radiation pressure is expected to affect the structure and efficiency of the accretion disc-corona system. This may cause departures from the typical nuclear properties of low-\ledd\ AGN, which have been largely studied so far.
  We present here the X-ray spectral analysis of 14 radio-quiet, $\lambda_\mathrm{Edd}\gtrsim1$ AGN at $0.4\leq z \leq 0.75$, observed with \emph{XMM-Newton}. Optical/UV data from simultaneous Optical Monitor observations have been also considered. These quasars have been selected to have relatively high values of black hole mass ($M_\mathrm{BH}\sim10^{8-8.5}M_\odot$) and bolometric luminosity ($L_\mathrm{bol} \sim 10^{46}$ erg s$^{-1}$), in order to complement previous studies of high-$\lambda_\mathrm{Edd}$ AGN at lower $M_\mathrm{BH}$ and $L_\mathrm{bol}$.  
  We studied the relation between \ledd\ and other key X-ray spectral parameters, such as the photon index ($\Gamma$) of the power-law continuum, the X-ray bolometric correction $k_\mathrm{bol,X}$ and the optical/UV-to-X-ray spectral index $\alpha_\mathrm{ox}$.
  Our analysis reveals that, despite the homogeneous optical and SMBH accretion properties, the X-ray properties of these high-\ledd\ AGN are quite heterogeneous.
  We indeed measured values of $\Gamma$ comprised between 1.3 and 2.5, at odds with the expectations based on previously reported $\Gamma-\lambda_\mathrm{Edd}$ relations, by which $\Gamma\geq2$ would be an ubiquitous hallmark of AGN with $\lambda_\mathrm{Edd}\sim1$. Interestingly, we found that $\sim30\%$ of the sources are X-ray weak, with an X-ray emission about a factor of $\sim10-80$ fainter than that of typical AGN at similar UV luminosities.
  The X-ray weakness seems to be intrinsic and not due to the presence of absorption along the line of sight to the nucleus.  This result may indicate that  high-\ledd\ AGN commonly undergo periods of intrinsic X-ray weakness. Furthermore, results from a follow-up monitoring with \emph{Swift} of one of these X-ray weak sources suggest that these periods can last for several years.}


  \maketitle


\section{Introduction}
\label{sec:intro}

Active galactic nuclei (AGN) are fuelled by accretion processes onto a central supermassive black hole (SMBH). 
The AGN accretion activity is usually quantified by the Eddington ratio, which is defined as the ratio between the AGN bolometric luminosity and the Eddington luminosity, i.e. $\lambda_\mathrm{Edd} = L_\mathrm{bol}/L_\mathrm{Edd}$.
The Eddington ratio is thus intimately connected with the accretion rate $\dot{M}$, since $\lambda_\mathrm{Edd} \propto \dot{M}/M_\mathrm{BH} \propto \dot{M}/\dot{M}_\mathrm{Edd}$, where $M_\mathrm{BH}$ is the black hole mass, $\dot{M}_\mathrm{Edd} = L_\mathrm{Edd}/(\eta c^2)$ and $\eta$ is the radiative efficiency. 

The bulk of the X-ray spectroscopic studies of AGN is focused on local ($z \lesssim 0.1$) sources with low-to-moderate ($ \lesssim0.3$) $\lambda_\mathrm{Edd}$ \citep[e.g.][]{nandra1994, piconcelli2005, bianchi2009, ricci2018}.
Because of their relative paucity in the local Universe \citep[e.g.][]{shankar2013, shirakata2019}, highly-accreting (i.e. $\lambda_\mathrm{Edd} \gtrsim 0.5$) AGN have been largely overlooked so far, with the notable exception of the Narrow Line Seyfert 1 galaxies \citep[NLSy1s; e.g.][]{brandt1997, gallo2006, costantini2007, jin2013, fabian2013, waddell2020}.
However NLSy1s represent a peculiar and restricted class of low-mass ($10^6-10^7\,M_\odot$) AGN with the narrowest permitted emission lines in type-1 AGN samples \citep[FWHM(H$\beta$) < 2000  km s$^{-1}$; e.g.][and references therein]{osterbrock1985, marziani2018}. 
In this sense, the NLSy1s may be a biased population, therefore it would be desirable to expand the study of the multi-wavelength properties of high-\ledd\ AGN to sources with Balmer lines widths of FWHM $\geq2000$ km s$^{-1}$ and hosting more massive SMBHs. Past and recent works support a close relation between the prominence of the optical Fe{\,\scriptsize II} emission and $\lambda_\mathrm{Edd}$ \citep[e.g.][]{marziani2001, shen2014, dou2016}, with the strong Fe{\,\scriptsize II} emission associated with high Eddington ratio. This result appears to be independent of the line width.

A clear understanding of the nuclear properties of high-$\lambda_\mathrm{Edd}$ AGN, regardless of their $L_\mathrm{bol}$
and $M_{\rm BH}$, would be extremely important for several reasons, as detailed as follows.
The standard model of accretion discs, developed by \citet{shakura1973}, predicts an optically thick, geometrically thin disc which is radiatively efficient. 
While this picture may hold for AGN in the low-$\lambda_\mathrm{Edd}$ regime, it is supposed to break down for high-$\lambda_\mathrm{Edd}$ sources, as the radiation pressure gains more importance with increasing accretion rate, and the circumnuclear region of the disc thickens vertically.
These systems are often called slim discs, as they are both optically- and geometrically-thick \citep[e.g.][]{abramowicz1988, chen2004, sadowski2011}.
Slim discs are also expected to have a low radiative efficiency because of the photon trapping effect.
The diffusion timescale of photons in optically- and geometrically-thick discs may be longer than the accretion timescale and they are advected into the central SMBH with the infalling material. This implies that the observed disc luminosity ($\sim L_\mathrm{bol}$) and, in turn, the \ledd\ are expected to saturate to a limiting value of approximately $5-10\,L_\mathrm{Edd}$ for steadily increasing accretion rates \citep[e.g.][]{mineshige2000}.
However, some critical issues on the theoretical side are still unsolved and the broadband spectral energy distributions of high- and low-$\lambda_\mathrm{Edd}$ AGN do not appear to exhibit clear differences \citep[e.g.][]{castello2017, liu2021}.

Furthermore, the study of high-$\lambda_\mathrm{Edd}$ AGN is important in terms of their cosmological implications. In fact, in the past few years there has been an increasing effort to detect quasars (i.e. AGN with $L_\mathrm{bol}$ $>$ 10$^{46}$ erg s$^{-1}$; QSOs hereafter) shining at $z\sim6-7$, when the universe was less than 1 Gyr old \citep[e.g.][]{Wu2015,banados2016,mazzucchelli2017}.
The vast majority of these QSOs typically host massive SMBHs (e.g. $M_\mathrm{BH}\geq10^9\,M_\odot$) and the process that allowed black hole seeds to grow up to billion solar masses in such a relatively short amount of time is still debated. 
Several scenarios have been proposed \citep[see, e.g., the review in][]{valiante2017} and it has been suggested that these SMBHs may evolve via gas accretion at a rate equal or above the Eddington rate.

{
\begin{table*}[ht]

\centering
\caption{{\it XMM-Newton} sample of high-$\lambda_\mathrm{Edd}$ QSOs and their general properties.} 
\renewcommand{\arraystretch}{1.5}
\begin{adjustbox}{max width=\textwidth}
\begin{threeparttable}
\begin{tabular}{c c c c c c c c c}
\hline\hline
Obj.\ ID & SDSS Name & RA & Dec & $z$ & $\log{(L_\mathrm{bol}/\mathrm{erg\,s}^{-1})}$ & $\log{(M_\mathrm{BH}/M_\odot)}$ & $\lambda_\mathrm{Edd}$ & $N_\mathrm{H,gal}$ \\
(1)&(2)&(3)&(4)&(5)&(6)&(7)&(8)&(9)\\
\hline
J0300$-$08 &   J030000.01$-$080356.9  &  03 00 00.01  &  $-$08 03 56.90   &  0.562  & 46.5  & 8.4  & 1.1  &  5.93   \\
J0809+46   &   J080908.13+461925.6    &  08 09 08.10  &  +46 19 25.30  &    0.657  &  46.5  & 8.5  & 0.9  &  3.60   \\
J0820+23   &   J082024.21+233450.4    &  08 20 24.20  &  +23 34 50.20  &    0.47  &   46.0  & 7.9  & 1.0  &  3.88   \\
J0940+46   &   J094033.75+462315.0    &  09 40 33.80  &  +46 23 15.00  &    0.696  &  46.4  & 8.3  & 1.1  &  1.06   \\
J1048+31   &   J104817.98+312905.8    &  10 48 18.00  &  +31 29 05.80   &    0.452  & 46.2  & 8.1  & 1.0  &  2.13   \\
J1103+41   &   J110312.93+414154.9    &  11 03 12.00  &  +41 41 55.00  &    0.402  &  46.4  & 8.3  & 1.0  &  0.81   \\
J1127+11   &   J112756.76+115427.1    &  11 27 56.80  &  +11 54 27.00  &    0.51  &   46.3  & 8.1  & 1.1  &  3.17   \\
J1127+64   &   J112757.41+644118.4    &  11 27 57.40  &  +64 41 18.20  &    0.695  &  46.4  & 8.2  & 1.2  &  0.79   \\
J1206+41   &   J120633.07+412536.1    &  12 06 33.10  &  +41 25 35.80  &    0.554  &  46.3  & 8.3  & 0.9  &  1.60   \\
J1207+15   &   J120734.62+150643.6    &  12 07 34.60  &  +15 06 43.60  &    0.75  &   46.5  & 8.4  & 0.9  &  2.26   \\
J1218+10   &   J121850.51+101554.1    &  12 18 50.50  &  +10 15 54.00  &    0.543  &  46.2  & 8.0  & 1.2  &  1.72   \\
J1245+33   &   J124511.25+335610.1    &  12 45 11.30  &  +33 56 10.10  &    0.711  &  46.6  & 8.5  & 1.1  &  1.20   \\
J1301+59   &   J130112.93+590206.7    &  13 01 12.93  &  +59 02 06.70   &    0.476  & 46.7  & 8.5  & 1.2  &  1.40   \\
J1336+17   &   J133602.01+172513.0    &  13 36 02.00  &  +17 25 13.10  &    0.552  &  46.6  & 8.5  & 0.9  &  1.64   \\
\hline
\end{tabular}
\begin{tablenotes}
\footnotesize\setlength\labelsep{0pt}
\item{\bf Notes:} (1) Abbreviated object ID; (2) SDSS IAU name; (3) right ascension (hours); (4) declination (degrees); (5) redshift from MS14; (6) bolometric luminosity (typical error of $<0.2$ dex) from the \citet{runnoe2012} relation; (7) black hole mass (typical error of $0.3-0.4$ dex; see e.g.\ \citealt{rakshit2020}) from the \citet{vestergaard2006} single-epoch virial relation; (8) Eddington ratio; (9) Galactic column density (10$^{20}$ cm$^{-2}$) from the full-sky H{\,\scriptsize{I}} map by \cite{bekhti2016}. 
\end{tablenotes}
\end{threeparttable}
\label{tab:AGNample}
\end{adjustbox}
\vspace{0.5cm}
\end{table*}
}

Finally, a high $\lambda_\mathrm{Edd}$ is invoked as a key ingredient to launch powerful nuclear outflows \citep[e.g.][]{proga2005, zubovas2013, king2015} which could be capable of controlling the growth and evolution of the host galaxy  \citep[e.g.][]{hopkins2010, fiore2017}.
These outflows are believed to deposit large amounts of energy and momentum into the interstellar medium \citep[e.g.][]{zubovas2012} and affect the host galaxy gas reservoir available for both star formation and SMBH accretion, offering a possible explanation for the observed $M_\mathrm{BH}-\sigma$ relation \citep[e.g.][]{ferrarese2000}. 
Accordingly, the AGN feedback mechanism should manifest itself in full force in high-$\lambda_\mathrm{Edd}$ sources, making them the ideal laboratory for probing the real impact of nuclear activity on the evolution of massive galaxies \citep[e.g.][]{reeves2009, nardini2015, tombesi2015, marziani2016, bischetti2017, bischetti2019, laurenti2021}.

The study of the X-ray properties of high-$\lambda_\mathrm{Edd}$ AGN allows to enlarge the dynamic range of the relations involving $\lambda_\mathrm{Edd}$ and X-ray spectral parameters and luminosity, by significantly populating the poorly-explored extreme end of the \ledd\ distribution and, thus, imposing crucial constraints on the strength of any possible correlation.
A long-standing issue concerning X-ray properties of high-\ledd\ AGN is the existence of a correlation between $\Gamma$ and \ledd, whereby a very steep (i.e. $\Gamma >$ 2) continuum is a prerogative of sources with $\lambda_\mathrm{Edd} \gtrsim 0.3$ \citep[e.g.][]{shemmer2008, risaliti2009, brightman2013, fanali2013, huang2020, liu2021}.
This correlation has been explained in terms of an enhanced UV emission from the accretion disc due to the high accretion rate, which leads to an increase of the radiative cooling of the X-ray corona (and a decrease of the electron temperature) and, in turn, to a steepening of the emerging X-ray continuum.
The discovery of such a strong dependence of $\Gamma$ on $\log\lambda_\mathrm{Edd}$, i.e. $\Gamma \sim 0.3\,\times\,\log{\lambda_\mathrm{Edd}} + 2$ \citep[e.g.][]{shemmer2008, risaliti2009, brightman2013}, has encountered an immediate interest for a possible application in X-ray extragalactic surveys, as it would allow to estimate the SMBH mass of AGN with reliable measurement of $\Gamma$, including type-2 AGN, for which commonly-used ``single-epoch'' $M_\mathrm{BH}$ estimators are not applicable.
However, \citet{trakhtenbrot2017} questioned the existence of a strong correlation between $\Gamma$ and $\lambda_\mathrm{Edd}$ reporting only a very weak correlation by analyzing a large sample of low-$z$ ($0.01 < z < 0.5$) sources with broadband X-ray spectra from the BAT AGN Spectroscopic Survey.

In addition, \citet{lusso2010} studied the hard X-ray bolometric correction, $k_\mathrm{bol,X} = L_\mathrm{bol} / L_\mathrm{2-10\,keV}$, and the optical/UV to X-ray spectral slope $\alpha_\mathrm{ox}$ of AGN in the COSMOS survey as a function of $\log\lambda_\mathrm{Edd}$.
They claimed for the existence of a linear relation between the logarithm of the Eddington ratio and the two X-ray quantities.
Although affected by large scatter, these relations lend support to a more physically-motivated scenario for the increase of $k_\mathrm{bol,X}$ as a function of $L_\mathrm{bol}$ \citep{marconi2004, martocchia2017, duras2020} which suggests an X-ray corona-accretion disc system with different properties in the most luminous (i.e. highly-accreting) AGN, producing a relatively weaker X-ray emission (as compared to the optical-UV emission from the accretion disc) than in “standard'' AGN.
A more refined description of the $k_\mathrm{bol,X}-\lambda_\mathrm{Edd}$ relation has been recently provided by \citet{duras2020}, who found $k_\mathrm{bol,X} \propto (\lambda_\mathrm{Edd})^{0.61}$ once a sizeable number of highly-accreting QSOs at $z\sim 2-4$ was included.
However, a full understanding of the quantitative behaviour of these relations, along with the $\Gamma-\lambda_\mathrm{Edd}$ one, and possible trends with $M_\mathrm{BH}$ or ${L_\mathrm{bol}}$ is still hampered by the lack of a large number of sources at $\lambda_\mathrm{Edd} \gtrsim 0.9$.
Indeed, the bulk of AGN that have been considered so far are accreting at low-to-moderate $\lambda_\mathrm{Edd}$ regimes ($\lesssim 0.3$).

To mitigate this bias, we present the X-ray spectroscopy of a sample of QSOs outside the local Universe ($z>0.1$) with $\lambda_\mathrm{Edd} \gtrsim 1$ and $M_\mathrm{BH}$ exceeding $\sim 10^8 M_\odot$.
This paper is organised as follows. We describe the general properties of our sample in Sect.\ \ref{sec:data}, as well as the {\it XMM-Newton} observations and data reduction. Spectral fits in the X-ray band and optical/UV photometry are detailed in Sect.\ \ref{sec:analysis}. We then present our results in Sect.\ \ref{sec:results}, which are discussed in Sect.\ \ref{sec:discussion}.
A $\Lambda$CDM cosmology ($H_0 = 67.7$ km s$^{-1}$ Mpc$^{-1}$, $\Omega_\mathrm{m} = 0.307$, $\Omega_\Lambda = 0.693$) from \emph{Planck} 2015 data is adopted throughout the paper \citep[][]{planck2016}.
All errors are quoted at $68\%$ confidence level \citep[$\Delta{C} = 1$;][]{avni1976, cash1979}.

\begin{table*}[ht]
\centering
\caption{Journal of the \emph{XMM-Newton} observations.}
\label{tab:obs}
\renewcommand{\arraystretch}{1.5}
\begin{adjustbox}{max width=\textwidth}
\begin{threeparttable}
\begin{tabular}{ c c c c c}
\hline\hline

Obj.\ ID & Obs.\ ID  & $t_\mathrm{net}$ &  $E|_\mathrm{h}$  & cts|$_\mathrm{bb}$ (cts|$_\mathrm{h}$)\\
(1)  & (2) & (3) & (4) & (5) \\  
\hline

J0300--08 &   0802220401     &  33.1/44.9/45.5 & 1.28 & 74/14/34 (24/4/9)        \\
J0809+46  &   0843830101     &  11.5/16.3/16.4 & 1.21 & 323/93/108 (187/59/73)      \\
J0820+23  &   0843830201     &  9.9/16.4/16.4  & 1.36 & 507/223/244 (114/63/74)     \\
J0940+46  &   0843830301     &  13.0/19.4/19.5 & 1.18 & 1831/505/469 (224/94/84)    \\
J1048+31  &   0843830401     &  8.5/16.4/16.4  & 1.38 & 392/183/203 (172/89/118)    \\
J1103+41  &   0843830501     &  13.7/16.8/16.8 & 1.43 & 999/269/270 (161/57/63)     \\
J1127+11  &   0843830601     &  2.5/12.5/13.8  & 1.32 & 380/504/568 (69/138/145)    \\
J1127+64  &   0843830701     &  16.5/24.7/24.7 & 1.18 & 51/20/30 (8/5/15)        \\
J1206+41  &   0843830801     &  4.1/10.9/11.7  & 1.29 & 230/114/112 (26/24/32)     \\
J1207+15  &   0843830901     &  10.8/16.4/16.4 & 1.14 & 948/331/321 (298/146/137)    \\
J1218+10  &   0843831001     &  4.1/18.7/17.8  & 1.30 & 449/340/445 (54/68/64)     \\
J1245+33  &   0843831101     &  12.1/16.1/16.1 & 1.17 & 58/15/13 (22/6/4)        \\
J1301+59  &   0304570101     &  8.5/11.2/12.1  & 1.36 & 4037/1336/1297 (696/337/339)  \\
J1336+17  &   0843831201     &  4.8/11.8/11.8  & 1.29 & 1801/1070/1205 (574/479/546)   \\
\hline
\end{tabular}

\begin{tablenotes}
\setlength\labelsep{0pt}\footnotesize
\item{\bf Notes:} (1) Abbreviated object ID; (2) \emph{XMM-Newton} observation ID; (3) net exposure for the pn/MOS1/MOS2 cameras (ks); (4) lower observer-frame bound of the hard X-ray energy range adopted to constrain the slope of the
underlying primary continuum, $E|_\mathrm{h}=2/(1+z)$ energy (keV); (5) net counts of the pn/MOS1/MOS2 cameras in the broadband (hard) $E=0.3-10$ keV ($E>2$ keV rest frame) observer-frame energy interval.  
\end{tablenotes}

\end{threeparttable}
\end{adjustbox}
\end{table*}


\section{Observations and data reduction}
\label{sec:data}

In this work we present the \emph{XMM-Newton} \citep[][]{jansen2001} observations of a sample of 14 AGN drawn from a larger sample of high-$\lambda_\mathrm{Edd}$ radio-quiet QSOs reported in \citet{marziani2014}, hereinafter MS14.
Specifically, they collected a sample of 43 radio-quiet QSOs in the range $z \sim 0.4-0.75$ with H$\beta$ coverage from a good-quality SDSS spectrum (see Table 1 in MS14). This ensures an accurate measure of $M_\mathrm{BH}$ and, in turn, $\lambda_\mathrm{Edd}$, being the full width at half maximum (FWHM) of the H$\beta$ emission line a more reliable single-epoch $M_\mathrm{BH}$ estimator than the FWHM of C{\,\scriptsize IV} \citep[e.g.][]{baskin2005, vietri2018, vietri2020}, which is typically used for QSOs at $z\gg1$.
The $M_\mathrm{BH}$ of these MS14 QSOs are in the range $10^{8-8.5} M_\odot$, while their $L_\mathrm{5100\,\AA}$-based $L_\mathrm{bol}$ are clustered around a few $10^{46}$ erg s$^{-1}$, with $\lambda_\mathrm{Edd}$ values spanning from $\sim 0.5$ to $\sim 2$. Such highly-accreting objects are also characterised by strong Fe{\,\scriptsize{II}} emission with $R_\mathrm{Fe{\,\scaleto{\mathrm{II}}{3pt}}} = \mathrm{EW(Fe{\,\scriptstyle II}}\,\lambda4570)/\mathrm{EW(H\beta)}$ $> 1$, and MS14 dubbed them as extreme Population A (xA) QSOs.
For our XMM-Newton observing program, we considered the 14 xA QSOs from the MS14 sample with $\lambda_\mathrm{Edd} \gtrsim 1$ and an estimated $L_\mathrm{bol} \geq 10^{46}$ erg s$^{-1}$. 

The main properties of our high-$\lambda_\mathrm{Edd}$ QSO sample are listed in Tab.\ \ref{tab:AGNample}. The bolometric luminosities are calculated by adopting the relation $\log{L_\mathrm{bol}} = 4.891 + 0.912 \log(\lambda\,L_\lambda)$, with $\lambda=5100\,$\AA, from \citet{runnoe2012}. 
The $M_\mathrm{BH}$ are estimated by using the single-epoch virial relation based on the FWHM of the H$\beta$ and the monochromatic luminosity at $\lambda=5100\,\AA$, as described by \citet{vestergaard2006}: $\log{M_\mathrm{BH}(\mathrm{H}\beta)} = \log \big[\mathrm{FWHM}/1000\,\mathrm{km\,s}^{-1} \big]^2 + \log\big[\lambda L_\lambda / 10^{44} \mathrm{erg\,s}^{-1} \big]^{0.5} + 6.91$. The virial mass estimates based on single-epoch spectra are typically dominated by the systematic uncertainty of  $\sim0.3-0.4$ dex \citep[e.g.][]{collin2006, shen2013, rakshit2020}, while the uncertainty associated with the \citet{runnoe2012} relation for $\log{L_\mathrm{bol}}$ is smaller than $0.2$ dex for the median value of $L_\mathrm{5100\,\AA}\sim7\times10^{41}$ erg s$^{-1}$ \AA$^{-1}$ of our sample. The errors on $\log M_\mathrm{BH}$ and $\log{L_\mathrm{bol}}$ then propagate to uncertainties in $\log{\lambda_\mathrm{Edd}}$. 

\emph{XMM-Newton} observed 12 out of the 14 high-\ledd\ AGN during the AO 18 Cycle campaign (PI: E.\ Piconcelli). The remaining two sources, namely J1301$+$59 and J0300$-$08, had publicly available \emph{XMM-Newton} observations at the time of our investigation. Detailed information on the \emph{XMM-Newton} data for each source in the sample are listed in Tab.\ \ref{tab:obs}.
Raw data from all sources have been retrieved from the \emph{XMM-Newton} Science Archive and then processed using the \emph{XMM-Newton} Science Analysis System (SAS v18.0.0) with the latest available calibration files. We take advantage of the full potential of \emph{XMM-Newton} in the X-ray energy interval $E=0.3-10$ keV (observer-frame) by collecting data by its primary instrument, i.e.\ the European Photon Imaging Camera \emph{EPIC}, equipped with the three X-ray CCD cameras, namely the pn \citep[][]{struder2001} and the two MOS detectors \citep[][]{turner2001}. Data reduction, filtering of high background periods and spectral extraction were performed according to the standard procedures described in the SAS web page\footnote{\url{https://www.cosmos.esa.int/web/xmm-newton/sas-threads}.}. For each object, we chose a circular region of radius $\sim10-20$ arcsec for the source extraction, and a nearby, larger source-free circular region for the background.
The spectra were binned to ensure at least one count per bin and modelled within the XSPEC v12.11.1c package by minimising the C-statistic \citep[][]{cash1979}. 

In addition, our study is complemented with the analysis of the simultaneous optical/UV data taken with the Optical Monitor \citep[OM;][]{mason2001}. Raw OM data are converted into science products by using the SAS task \texttt{omichain}. We use the task \texttt{om2pha} to convert the OM photometric points into a suitable format for XSPEC \citep[][]{arnaud1996}.

\section{Data analysis}
\label{sec:analysis}
\subsection{X-ray spectroscopy}
\label{sec:spec}

We adopted the following procedure to analyse the X-ray spectrum of each source in our sample. Data from \textit{EPIC}-pn, MOS1 and MOS2 were always considered simultaneously and fitted together, taking into account an intercalibration constant ($<10\%$) between the three instruments. We first ignored all the data outside the $E=0.3-10$ keV observer-frame energy interval. Initially, only data in the intrinsic hard X-ray range, i.e. with $E>2$ keV in the rest frame of the source, corresponding to $E>2/(1+z)$ keV in the observer frame, were considered.
This hard X-ray portion of the spectrum was modelled with a power law modified by Galactic absorption to constrain the slope of the underlying primary continuum without any contamination from the likely soft excess component emerging at lower energies, which may strongly affect the measurement of the photon index $\Gamma$. From this simple spectral fit, we found the bulk of $\Gamma$ values to be comprised between $\sim 1.7$ and $2.3$, which is the typical range of radio-quiet AGN \citep[e.g.][]{piconcelli2005, gliozzi2020}. A flatter hard-band continuum (i.e. $\Gamma \sim 1.2-1.3$) was found for two sources, namely J1048+31 and J1245+33, although the latter was affected by large uncertainty. Finally, the $\Gamma$ value of J0300$-$08 resulted basically unconstrained due to the limited statistics.

{
\begin{table*}[ht]

\centering
\caption{Best-fitting spectral parameters for the broadband ($E=0.3-10$ keV) X-ray spectra of each sample source, measured by assuming a model consisting of a power law plus a blackbody component, modified by Galactic absorption.} 
\label{tab:spectral_analysis}
\renewcommand{\arraystretch}{1.5}
\begin{adjustbox}{max width=\textwidth}
\begin{threeparttable}
\begin{tabular}{ c c c c c c c c c c }
\hline\hline

Obj.\ ID & $\Gamma$ & $kT_\mathrm{bb}$ & (C-stat/d.o.f.) (d.o.f.) & $F_{0.5-2\,\mathrm{keV}}$  & $F_{2-10\,\mathrm{keV}}$ & $L_{0.5-2\,\mathrm{keV}}$  & $L_{2-10\,\mathrm{keV}}$  & $k_\mathrm{bol,X}$ & $R_\mathrm{S/P}$\\

 (1)  & (2) & (3) & (4) & (5) & (6)  & (7) & (8) & (9) & (10) \\
\hline
J0300--08 &  1.9 (fixed)  & $140 \pm 30$      & 1.22 (60) & $3.7_{-0.7}^{+0.3} \times10^{-15}$  & $4.3_{-0.5}^{+0.9} \times10^{-15}$  & $9.4_{-0.7}^{+0.8}\times10^{42}$  &  $6_{-1}^{+1}\times10^{42}$  & $5100\pm2500$         & 1.24 \\
J0809+46$^a$ &  $2.10 \pm 0.06$ & --              & 1.16 (169) & $2.8_{-0.2}^{+0.1} \times 10^{-14}$ & $8.6_{-0.9}^{+0.7} \times 10^{-14}$ & $2.4_{-0.4}^{+0.3} \times 10^{44}$  &  $2.3_{-0.4}^{+0.2} \times 10^{44}$ & $150\pm70$     &  --  \\
J0820+23 &  $2.26 \pm 0.09$ & $80 \pm 20$         & 1.08 (173) & $8.3_{-0.6}^{+0.3} \times 10^{-14}$ & $7.1_{-0.6}^{+0.5} \times 10^{-14}$ & $1.13_{-0.07}^{+0.07} \times 10^{44}$ &  $6.9_{-0.5}^{+0.5} \times 10^{43}$ & $150\pm70$      & 0.28 \\
J0940+46 &  $2.3 \pm 0.1$ & $134 \pm 4$         & 0.90 (205) & $1.18_{-0.03}^{+0.03}\times 10^{-13}$& $5.3_{-0.6}^{+0.5} \times 10^{-14}$ & $6.7_{-0.2}^{+0.2} \times 10^{44}$ &  $1.4_{-0.2}^{+0.2} \times 10^{44}$ & $200\pm90$        & 2.59 \\
J1048+31$^b$ &  $1.27 \pm 0.08$ & $120 \pm 20$      & 0.81 (210) & $5.1_{-0.4}^{+0.2} \times 10^{-14}$ & $1.8_{-0.2}^{+0.2} \times 10^{-13}$& $4.4_{-0.3}^{+0.3} \times 10^{43}$ &  $1.1_{-0.1}^{+0.1} \times 10^{44}$ & $140\pm70$        & 0.55 \\
J1103+41 &  $1.6 \pm 0.1$ & $122 \pm 6$         & 0.84 (196) & $6.9_{-0.2}^{+0.2} \times 10^{-14}$ & $7.7_{-0.9}^{+0.7} \times 10^{-14}$ & $6.9_{-0.3}^{+0.3} \times 10^{43}$ &  $4.1_{-0.5}^{+0.6} \times 10^{43}$ & $600\pm300$       & 2.42 \\
J1127+11 &  $2.3 \pm 0.1$ & $170 \pm 20$        & 0.91 (192) & $2.24_{-0.06}^{+0.05}\times 10^{-13}$& $1.4_{-0.2}^{+0.2} \times 10^{-13}$ & $3.6_{-0.3}^{+0.4} \times 10^{44}$ &  $1.7_{-0.3}^{+0.3} \times 10^{44}$ & $100\pm50$        & 0.52 \\
J1127+64 &  $1.5 \pm 0.4$ & $110 \pm 20$     & 0.77 (53) & $3.8_{-1.3}^{+0.2} \times 10^{-15}$ & $7_{-4}^{+2} \times 10^{-15}$    & $2.4_{-0.7}^{+0.3} \times 10^{43}$ &  $1.2_{-0.5}^{+0.7} \times 10^{43}$ & $2000_{-1300}^{+1500}$  & 4.37 \\
J1206+41 &  $1.4 \pm 0.2$ & $128 \pm 7$        & 0.70 (97) & $6.2_{-0.5}^{+0.2} \times 10^{-14}$ & $8_{-2}^{+2} \times 10^{-14}$    & $1.90_{-0.07}^{+0.09} \times 10^{44}$ &  $8_{-2}^{+2} \times 10^{43}$ & $300\pm100$         & 5.52 \\
J1207+15 &  $2.0 \pm 0.1$ & $160^{+30}_{-50}$      & 0.97 (224) & $1.04_{-0.03}^{+0.02}\times 10^{-13}$& $1.1_{-0.1}^{+0.1} \times 10^{-13}$ & $3.9_{-0.4}^{+0.4} \times 10^{44}$ &  $3.1_{-0.4}^{+0.5} \times 10^{44}$ & $100\pm50$        & 0.45 \\
J1218+10 &  $2.5 \pm 0.2$ & $160 \pm 10$        & 0.98 (162) & $1.43_{-0.04}^{+0.05}\times 10^{-13}$& $5.1_{-0.8}^{+0.6} \times 10^{-14}$ & $3.1_{-0.3}^{+0.3} \times 10^{44}$ &  $8_{-2}^{+2} \times 10^{43}$ & $200\pm100$          & 0.97 \\
J1245+33 &  1.3 (fixed) & $120 \pm 20$      & 0.96 (44) & $5.2_{-1.0}^{+0.4} \times 10^{-15}$ & $1.3_{-0.2}^{+0.2} \times 10^{-14}$ & $2.6_{-0.1}^{+0.3} \times 10^{43}$ &  $2.1_{-0.4}^{+0.4} \times 10^{43}$ & $2000\pm1000$         & 2.93 \\
J1301+59 &  $2.01 \pm 0.05$ & $122 \pm 3$      & 0.98 (295) & $4.65_{-0.09}^{+0.06}\times10^{-13}$ & $3.9_{-0.2}^{+0.2}\times10^{-13}$  & $7.0_{-0.2}^{+0.2}\times10^{44}$  &  $3.6_{-0.2}^{+0.2}\times10^{44}$  & $140\pm60$        & 1.25 \\
J1336+17 &  $1.86 \pm 0.04$ & $100^{+40}_{-30}$ & 1.01 (336) & $4.3_{-0.1}^{+0.1} \times 10^{-13}$ & $6.4_{-0.2}^{+0.2} \times 10^{-13}$ & $6.0_{-0.2}^{+0.2} \times 10^{44}$ &  $7.8_{-0.3}^{+0.3} \times 10^{44}$ & $50\pm20$         & 0.10 \\

\hline
\end{tabular}

\begin{tablenotes}
\setlength\labelsep{0pt}
\item\textbf{Notes:} (1) Abbreviated object ID; (2) photon index of the power-law continuum component; (3) rest-frame temperature of the blackbody component  (eV); (4) ratio between the C-stat value of the spectral fit and the degrees of freedom; (5) flux in the $0.5-2$ keV  band (erg cm$^{-2}$ s$^{-1}$); (6) flux in the  $2-10$ keV band (erg cm$^{-2}$ s$^{-1}$); (7) luminosity in the $0.5-2$ keV band (erg s$^{-1}$); (8) luminosity in the $2-10$ keV band (erg s$^{-1}$); (9)  X-ray bolometric correction; (10)  value of the $R_\mathrm{S/P}$ parameter. 
$^a$ the best-fit model of J0809$+$46 is a power law modified by a warm absorber component (see Sect.\ \ref{sec:spec} and Fig.\ \ref{fig:spectra} of Appendix \ref{sec:appA}) and it does not require a blackbody component.
$^b$ the best-fit model of J1048$+$31 also includes a Gaussian emission line component, see Fig.\ \ref{fig:spectra} of Appendix \ref{sec:appA}.
\end{tablenotes}

\end{threeparttable}
\end{adjustbox}

\end{table*}
}

{
\begin{table*}[ht]

\centering
\caption{Optical/UV properties of the QSOs. Monochromatic luminosities, corrected for both Galactic and local dust extinction, are reported in units of $10^{30}$ erg s$^{-1}$ Hz$^{-1}$. }
\label{tab:optical_data}
\renewcommand{\arraystretch}{1.5}
\begin{adjustbox}{max width=\textwidth}
\begin{threeparttable}
\begin{tabular}{ c c c c c c c c c }
\hline\hline

Obj.\ ID & E$(B-V)_\mathrm{int}$ & $L_\mathrm{B}$ & $L_\mathrm{U}$ & $L_\mathrm{UVW1}$ & $L_\mathrm{UVM2}$ & $L_\mathrm{UVW2}$ & $L_\mathrm{2500\,\AA}$ & $\alpha_\mathrm{ox}$\\
(1)  & (2) & (3) & (4) & (5) & (6)  & (7) & (8) & (9) \\
\hline

J0300--08& $<0.01$   &  --           &  $5.68\pm0.05$     & $5.54\pm0.04$     & $4.6\pm0.1$       & $4.2\pm0.1$   & $5.78\pm0.05$  & $-2.28\pm0.03$   \\
J0809+46 & $<0.01$   &  $14.6 \pm 0.1$     &  $13.7\pm0.1$      & $14.0\pm0.1$      & --           & --        & $14.3\pm0.1 $  & $-1.78\pm0.02$   \\
J0820+23 & $<0.01$   &  $1.91 \pm 0.06$     &  $1.57\pm0.04$     & $1.61\pm0.03$     & --           & --        & $1.65\pm0.04$  & $-1.60\pm0.01$   \\   
J0940+46 & 0.06 &  $4.82 \pm 0.09$     &  $4.11\pm0.06$     & $4.02\pm0.07$     & --           & --        & $4.66\pm0.08$  & $-1.65\pm0.02$   \\   
J1048+31 & 0.20 &  $2.54 \pm 0.08$     &  $1.87\pm0.06$     & $1.73\pm0.06$     & --           & --        & $1.99\pm0.05$  & $-1.71\pm0.02$   \\   
J1103+41 & 0.05 &  $4.89 \pm 0.04$     &  $4.29\pm0.03$     & $3.80\pm0.04$     & --           & --        & $4.33\pm0.03$  & $-1.94\pm0.02$   \\   
J1127+11 & $<0.01$   &  --           &  $2.47\pm0.03$     & $2.41\pm0.04$     & $2.33\pm0.06$      & --        & $2.51\pm0.03$  & $-1.51\pm0.02$   \\   
J1127+64 & $<0.01$   &  $4.14 \pm 0.07$     &  $3.49\pm0.05$     & $3.39\pm0.05$     & --           & --        & $3.99\pm0.06$  & $-2.16\pm0.04$   \\
J1206+41 & 0.08 &  --           &  $1.79\pm0.04$     & $1.52\pm0.03$     & --           & --        & $2.01\pm0.05$  & $-1.73\pm0.04$   \\   
J1207+15 & 0.13 &  $5.2 \pm 0.2$      &  $4.7\pm0.1$      & $3.8\pm0.1$      & --           & --        & $5.1\pm0.2$   & $-1.57\pm0.02$   \\   
J1218+10 & $<0.01$   &  $5.38 \pm 0.07$     &  --	          & $3.45\pm0.04$     & --           & --        & $4.60\pm0.05$  & $-1.71\pm0.03$   \\   
J1245+33 & $<0.01$   &  $10.5 \pm 0.1$     &  $10.32\pm0.09$     & $9.9\pm0.1$      & --           & --        & $10.5\pm0.1$   & $-2.25\pm0.04$   \\   
J1301+59 & $<0.01$   &  --           &  $12.51\pm0.03$     & $12.34\pm0.04$     & $12.17\pm0.06$     & --        & $12.58\pm0.03$  & $-1.70\pm0.01$   \\   
J1336+17 & $<0.01$   &  $10.1 \pm 0.1$     &  $9.02\pm0.08$     & $8.9\pm0.1$      & --           & --        & $9.49\pm0.09$  & $-1.55\pm0.01$   \\  
\hline

\end{tabular}

\begin{tablenotes}
\setlength\labelsep{0pt}\footnotesize
\item\textbf{Notes:} (1) Abbreviated object ID; (2) internal reddening; (3) $B$ rest-frame luminosity; (4) $U$ rest-frame luminosity; (5) $UVW1$ rest-frame luminosity; (6) $UVM2$ rest-frame luminosity; (7) $UVW2$ rest-frame luminosity; (8) $2500\,\AA$ rest-frame luminosity; (9) $\alpha_\mathrm{ox}$. 
\end{tablenotes}

\end{threeparttable}

\end{adjustbox}
\vspace{0.5cm}
\end{table*}
}

We then extended our spectral analysis to the whole $E=0.3-10$ keV energy interval by including the soft X-ray portion. The extrapolation of the model to soft energies showed for all but one sources (J0809$+$46)\footnote{A detailed spectral analysis of J0809+46 will be presented in a forthcoming paper (Piconcelli et al., \emph{in prep.}). The X-ray spectrum of this source shows no soft excess: at low energies its spectrum is dominated by an ionised absorption component with high column density (see Fig. \ref{fig:spectra}).} the presence of a soft excess component in their X-ray spectra and we tried to model this feature by considering two different spectral components, i.e., a blackbody or an additional power law, both modified by Galactic absorption. Clearly, we are aware that both models merely provide a phenomenological explanation of the soft excess, whose physical origin is still debated \citep[e.g.][]{sobolewska2007, fukumura2016, petrucci2018, middei2020}.
However, the blackbody model provided the best description of the soft excess in terms of C-statistic for all the spectra. 

From our broadband spectral analysis we obtained the best-fit values of the photon index $\Gamma$ and the blackbody temperature $kT_\mathrm{bb}$ as listed in Tab.\ \ref{tab:spectral_analysis}. We find that our high-\ledd\ AGN do have $kT_\mathrm{bb}$ which are consistent with those typically measured for type-1 AGN \citep[e.g.][]{piconcelli2005, bianchi2009}. For two sources, J1245$+$33 and J0300$-$08, we fixed the photon index of the power-law component due to the limited quality of spectral data. Specifically, for J1245+33 we fixed $\Gamma=1.3$, which is the value derived from power-law fit to the hard X-ray portion of the spectrum, while for J0300$-$08 we adopted the canonical value of $\Gamma$ = 1.9 found for radio-quiet quasars in previous works \citep[e.g.][]{reeves2000, piconcelli2005}, since the photon index was largely unconstrained in this case. 
No significant evidence for Fe K$\alpha$ emission was found in the spectra, except for J1048+31, for which we report the presence of an emission line at $E=6.52_{-0.04}^{+0.05}$ keV with a rest-frame equivalent width of $EW = 400_{-140}^{+150}$ eV (see Fig.\ \ref{fig:spectra} of Appendix \ref{sec:appA}). 
\begin{figure}[t]
\centering
\vspace{-0.43cm}
\includegraphics[width=0.95\columnwidth]{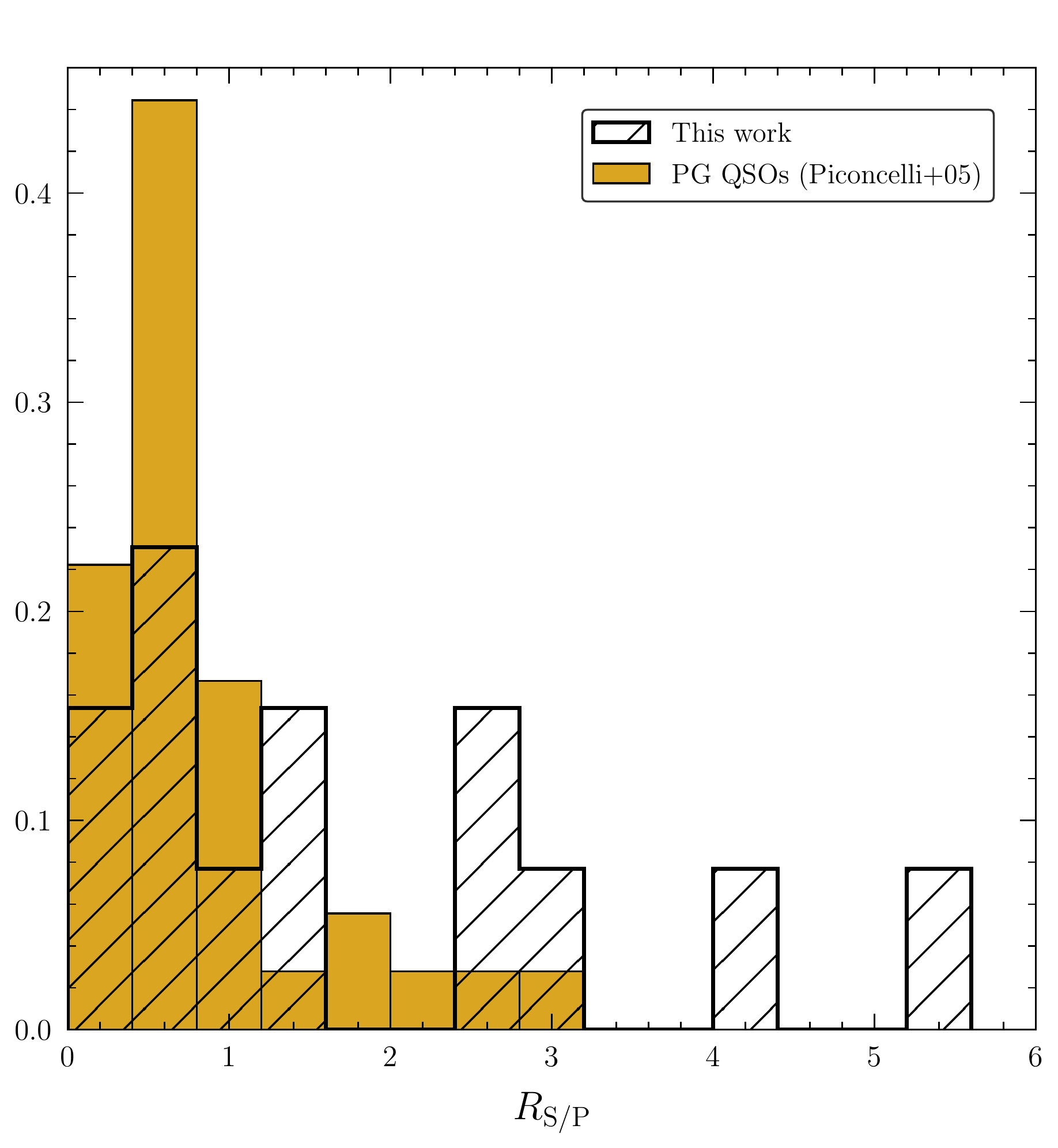}
\caption{Distribution of $R_\mathrm{S/P} = L^\mathrm{0.5-2\,\mathrm{keV}}_\mathrm{BB}/L^\mathrm{0.5-2\,\mathrm{keV}}_\mathrm{PL}$ for our AGN sample (in black) and the sample of PG QSOs from \cite{piconcelli2005} (in gold). $R_\mathrm{S/P}$ measures the relative strength between the luminosity of the blackbody and power law component in the $E=0.5-2$ keV energy band.}
\label{fig:Rsp}
\end{figure}
Also shown in Fig.\ \ref{fig:spectra} are the X-ray spectra, and the corresponding best fit, of all the 14 high-\ledd\ AGN analysed in this work.
Moreover, in Appendix \ref{sec:serendipitous} (see also Fig. \ref{fig:serendipitous}) we present the results of the X-ray spectral analysis of some interesting serendipitous sources located in the \emph{XMM-Newton} field of view of our targeted high-$\lambda_\mathrm{Edd}$ AGN. 
Table \ref{tab:spectral_analysis} also lists the values of the flux and luminosity of the 14 high-\ledd\ QSOs in both the soft ($E=0.5-2$ keV) and the hard X-ray ($E=2-10$ keV) band, as well as the X-ray bolometric correction $k_\mathrm{bol,X}$ resulting from this broadband spectral fit. 
Furthermore, we also calculated the parameter $R_\mathrm{S/P} = L^\mathrm{0.5-2\,\mathrm{keV}}_\mathrm{BB}/L^\mathrm{0.5-2\,\mathrm{keV}}_\mathrm{PL}$ which is a proxy of the relative strength between the luminosity of the blackbody and power law components (i.e. the soft excess and the primary continuum, respectively) in the $E=0.5-2$ keV observer-frame energy band. Figure \ref{fig:Rsp} shows the distribution of $R_\mathrm{S/P}$ for our sample of high-$\lambda_\mathrm{Edd}$ AGN compared to that derived by \citet{piconcelli2005} for a sample of Palomar-Green (PG) QSOs with $-1.2 \leq \log{\lambda_\mathrm{Edd}} \leq 0.7$. Such a sample of PG QSOs contains 38 bright objects\footnote{We do not consider PG 1226+023, aka 3C 273, and PG 1115+080: the former is a well-known blazar-line AGN, while the latter is a lensed QSO at redshift $z\sim1.7$.} with an absolute magnitude $M_B < -23$ in the Johnson's $B$ band, and with redshift $z\lesssim1.5$, which is broadly comparable to that of our high-\ledd\ AGN. 
It is worth noting that the distribution of $R_\mathrm{S/P}$ for the PG QSOs does not extend to values larger than $\sim 3$ and is largely dominated by sources with $R_\mathrm{S/P}$ $<$ 1, while for our sample the distribution appears smoother and much broader, by reaching values of $R_\mathrm{S/P}\sim6$.

\subsection{Optical/UV photometry}
\label{sec:om}

Each source in the sample has at least two available measurements of optical/UV photometry taken with the OM, which are simultaneous to the X-ray data, with one always collected in the \textit{UVW1} (2910 \AA) filter. 

The other photometric points are distributed among the \textit{B} (4500 \AA), \textit{U} (3400 \AA), \textit{UVM2} (2310 \AA), and \textit{UVW2} (2120 \AA) filters.
The extinction-corrected optical/UV fluxes are derived as follows. 
We first consider the optical spectrum of each source from the Sloan Digital Sky Survey \citep[SDSS;][]{york2000} Legacy campaign.
These spectra are the same used by MS14 and were downloaded from the SDSS SkyServer web tool\footnote{\url{http://skyserver.sdss.org/dr16/en/home.aspx}.}.
Galactic extinction is taken into account by considering the color excess measurement from the reddening map of \citet{schlafly2011}, provided by the NASA/IPAC Infrared Science Archive (IRSA) website\footnote{\url{https://irsa.ipac.caltech.edu/applications/DUST/}.}.
The spectra are then corrected for Galactic extinction according to the Milky Way reddening law of \citet{fitzpatrick1999}, assuming $R_\mathrm{V} = 3.1$.

We estimate the possible contribution of the host galaxy starlight to the UV/optical emission by modelling the optical spectrum as a combination of AGN and galaxy components, following the procedure described in \citet{vagnetti2013}. We find that the host galaxy contribution is completely negligible for $\sim80\%$ of the sample. For the remaining sources (i.e.\ J1048+31, J1206+41 and J1218+10), this host galaxy contribution results only to be marginal ($\lesssim10\%$) and can be ignored.

\begin{figure}[ht]
\includegraphics[width=\columnwidth]{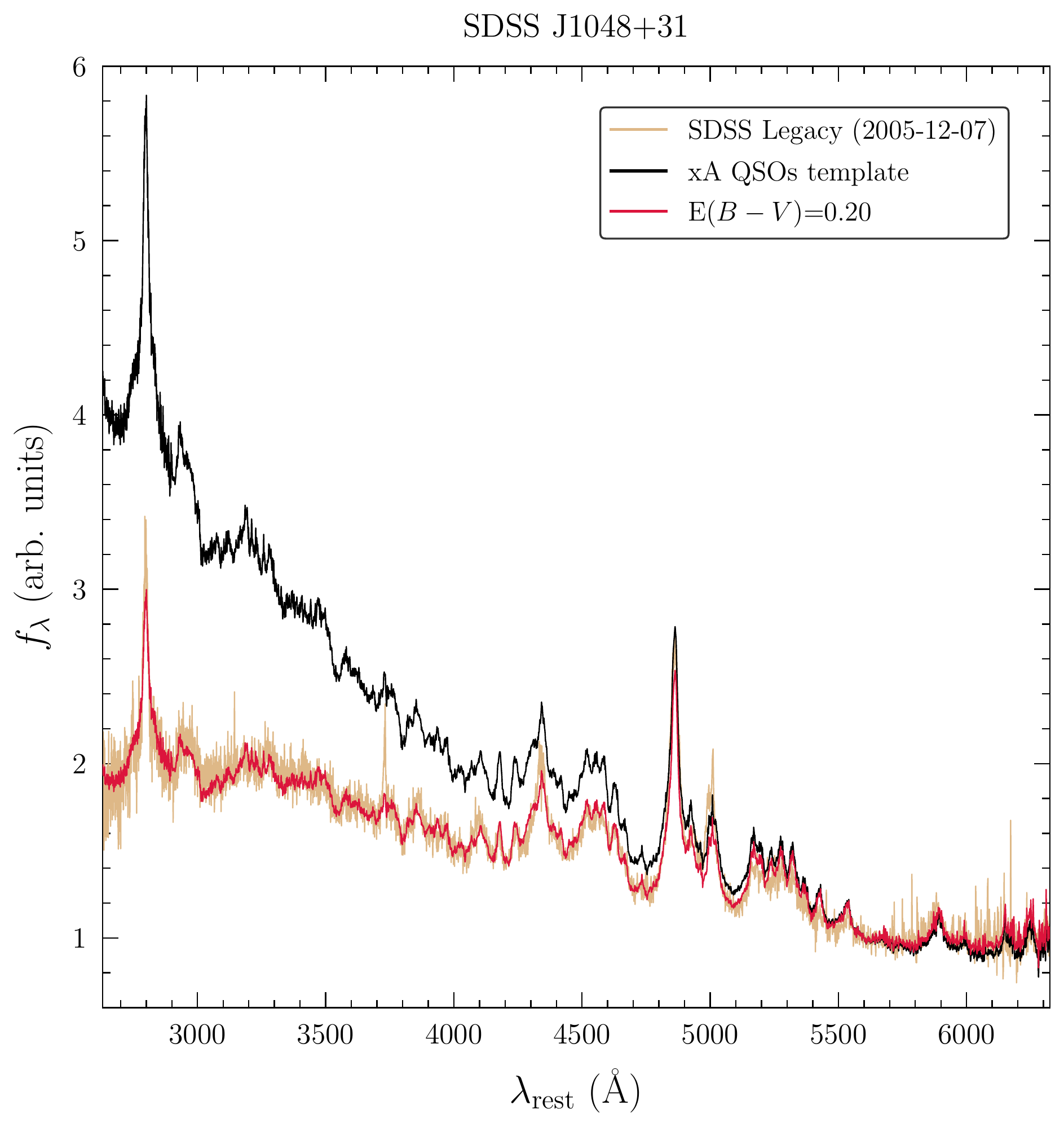}
\caption{SDSS spectrum of J1048+31 at $z=0.452$ (in gold) compared to the composite template of xA QSOs created by Marziani et al.\ (\emph{in prep.}).
The template (in black) is extinguished according to the SMC extinction law by \citet{prevot1984}, with progressively increasing values of E$(B-V)$. The value of E$(B-V)$ for which the reddened template (in red) matches with the observed SDSS spectrum, corresponds to our estimate of the internal reddening of the source. J1048+31 has the largest E$(B-V)$ among our sources, i.e. E$(B-V)$ = 0.2 }
\label{fig:reddening}
\end{figure}

Furthermore, we look for indications of internal reddening by comparing the SDSS spectra with two different templates: (i) the SDSS composite QSO spectrum by \citet{vandenberk2001}; and (ii) a median composite spectrum of xA QSOs built from the full sample of highly-accreting QSOs in MS14 (see also Marziani et al., \emph{in prep.}). For each source, the SDSS spectrum and the templates are normalised to unit flux at rest-frame wavelengths corresponding to $\lambda_\mathrm{obs}\geq8000\,\AA$ in the observer frame. 
We then apply the Small Magellanic Cloud (SMC) extinction law by \citet{prevot1984} to both templates as shown in Fig.\ \ref{fig:reddening}. The choice of an SMC-like extinction law stems from its capability of reproducing dust reddening of QSOs at all redshifts \citep[e.g.][]{richards2003, hopkins2004, bongiorno2007, gallerani2010, krawczyk2015}. Although the two templates yield similar results, we adopt the xA QSOs template since it provides a more accurate description of the optical/UV behaviour of our sources.
Even if the majority of our QSOs do not suffer from significant internal reddening, for a sizeable fraction of the sample ($\sim35\%$), the estimated E$(B-V)$ is $\geq0.05$ and two sources, namely J1048+31 and J1207+15, show an E$(B-V)$ of $0.13$ and $0.2$, respectively (see Tab.\ \ref{tab:optical_data}). The typical uncertainty on these reddening estimates is $\sim0.01$.
OM photometric data are corrected for the effects of both Galactic and internal dust extinction according to the same procedure.

Finally, the value of the monochromatic UV luminosity $L_\mathrm{2500\,\AA}$ for each QSO is obtained by linearly interpolating (or, if necessary, extrapolating) the OM data points at rest-frame 2500\,\AA. This is used jointly with the monochromatic X-ray luminosity at 2 keV estimated from the X-ray spectra (see Sect.\ \ref{sec:spec}) to calculate $\alpha_\mathrm{ox}$ as:
\begin{equation}
\begin{split}
  \alpha_\mathrm{ox} &= \log{\Bigg[\frac{L_\nu(2\,\mathrm{keV})}{L_\nu(2500\,\AA)}\Bigg]}\,\Bigg/\,\log{\Bigg[\frac{\nu(2\,\mathrm{keV})}{\nu(2500\,\AA)}\Bigg]} \\
  & \simeq 0.384\,\log{\Bigg[\frac{L_\nu(2\,\mathrm{keV})}{L_\nu(2500\,\AA)}\Bigg]}\,.
\end{split} 
\end{equation}

\noindent We find that in our high-$\lambda_\mathrm{Edd}$ AGN sample the bulk of the $\alpha_\mathrm{ox}$ values is around $-1.8$, with a minimum value of approximately $-2.3$.

\begin{figure}[t]
\includegraphics[width=\columnwidth]{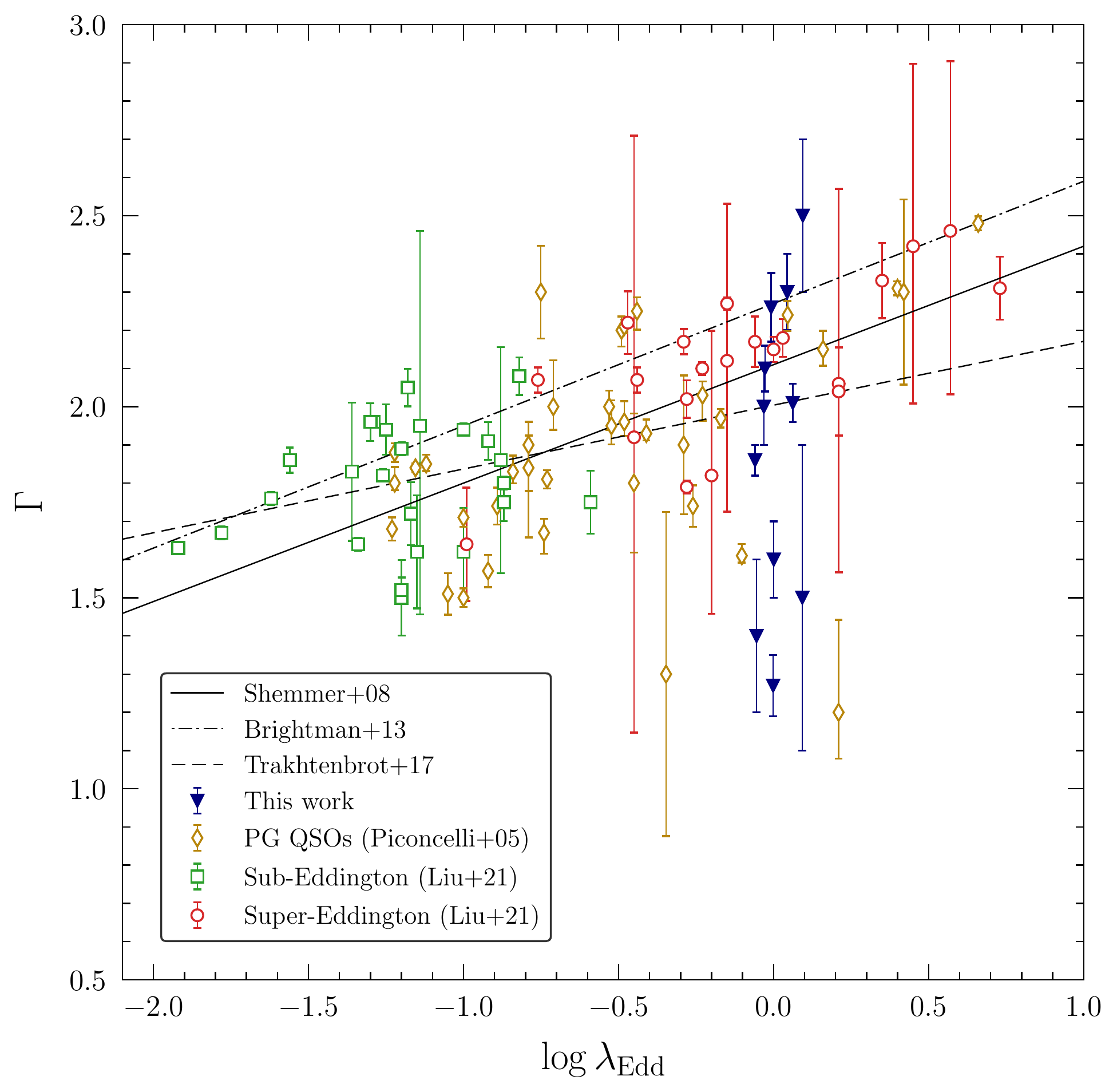}
\caption{$\Gamma$ vs $\log{\lambda_\mathrm{Edd}}$ for the high-\ledd\ QSOs in our sample (blue triangles). Two sources, J0300--08 and J1245+33, are excluded since their $\Gamma$ were unconstrained by \emph{XMM-Newton} data. J0940+46 and J1127+11 are described by the same point, since they share the same values of $\Gamma$ and $\lambda_\mathrm{Edd}$. Black solid, dashed and dash-dotted lines indicate the expected relation from other studies. Gold diamonds represent the sample of PG QSOs from \cite{piconcelli2005}. Green squares and red circles indicate the sub- and super-Eddington (i.e. $\lambda_\mathrm{Edd}$ $\geq$ 0.3) AGN from \citet{liu2021}, respectively.}
\label{fig:gammaVSedd}
\end{figure}

\section{Results}
\label{sec:results}

\subsection{X-ray continuum slope}

\begin{figure}[t]
\includegraphics[width=\columnwidth]{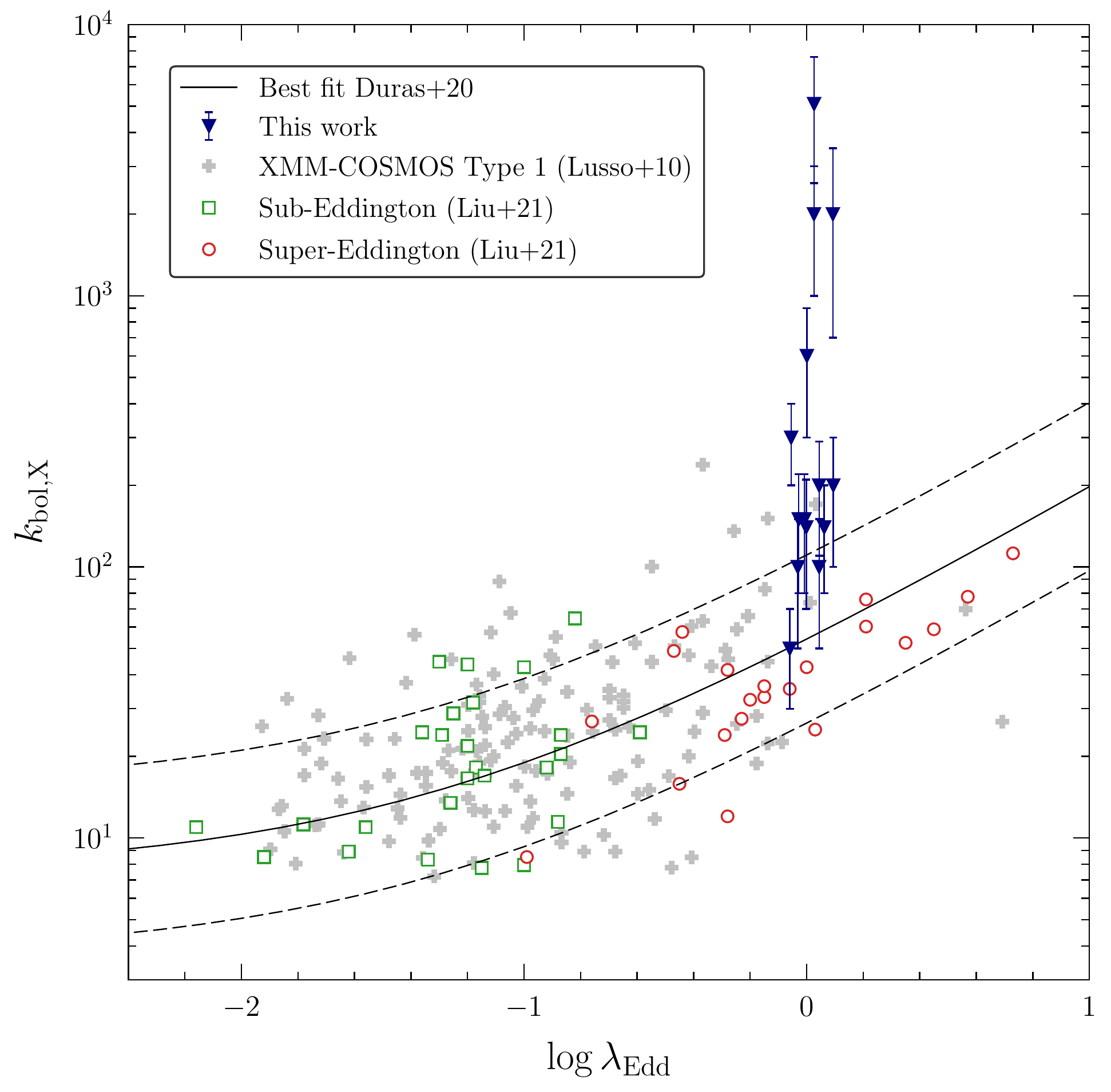}
\caption{$k_\mathrm{bol,X}$ vs $\log{\lambda_\mathrm{Edd}}$ for the high-$\lambda_\mathrm{Edd}$ AGN in our sample (blue triangles) compared to the XMM-COSMOS type-1 AGN (light grey crosses) of \citet{lusso2010} with available measurement of Eddington ratio. Black solid line refers to the best-fit relation from \citet{duras2020}. Black dashed lines describe the $1\sigma$ spread of the above relation. Green squares and red circles indicate the sub- and super-Eddington AGN from \citet{liu2021}, respectively.}
\label{fig:kbolVSlbol}
\end{figure}

In Sect.\ \ref{sec:data} we mentioned that our AGN have been selected to span a narrow interval of $\lambda_\mathrm{Edd}$, with $0.9 \leq \lambda_\mathrm{Edd} \leq 1.2$. This enables us to investigate the properties of these sources on a statistical sound basis and  reduce the effect of washing out possible correlations due to considering sources in a wide range of $\lambda_\mathrm{Edd}$.
Nonetheless, Fig.\ \ref{fig:gammaVSedd} shows that the continuum slope of the high-$\lambda_\mathrm{Edd}$ AGN exhibits quite a large scatter.
The values of $\Gamma$ found by our analysis indeed span from $\sim1.3$ to $\sim2.5$, with a non-negligible number of sources having quite flat photon indices $\Gamma \lesssim 1.6$.
Furthermore, the flat X-ray continuum does not seem to be exclusively associated with an X-ray weak state (see Sect.\ \ref{sub:weak}). Indeed, out of the four $\Gamma\leq1.6$ sources in Fig.\ \ref{fig:gammaVSedd}, only two are also X-ray weak (namely, J1103+41 and J1127+64).
Interestingly, the presence of a consistent fraction ($\sim30\%$) of high-\ledd\ AGN with flat $\Gamma$ in our sample is at odds with the expectations based on the popular $\Gamma - \lambda_\mathrm{Edd}$ relations reported in previous works \citep[e.g.][]{shemmer2008, risaliti2009, brightman2013}, which predict steeper $\Gamma$ of $\sim 2-2.3$ for AGN with $\lambda_\mathrm{Edd} \sim 1$. 

However,  \citet{trakhtenbrot2017} found that the above relation between $\Gamma$ and \ledd\ could be less steep than expected. 
They considered 228 hard X-ray selected AGN from the \emph{Swift}/BAT AGN spectroscopic survey (BASS), with redshift $0.01 < z < 0.5$, 
$0.001 \lesssim \lambda_\mathrm{Edd} \lesssim 1$, and $6 \lesssim \log{(M_\mathrm{BH}/M_\odot)} \lesssim 9.5$, which benefit from high-quality and broadband X-ray spectral coverage.
For their BASS sample, \citet{trakhtenbrot2017} found a statistically significant, albeit very weak correlation between $\Gamma$ and $\lambda_\mathrm{Edd}$.
To complete the picture, in a recent paper, \citet{liu2021} analysed a sample of 47 local, radio-quiet AGN with an accurate $M_\mathrm{BH}$ measurement from reverberation mapping, which includes both sub- and super-Eddington (i.e. $\lambda_\mathrm{Edd}$ $\geq$ 0.3) sources.
They found a significant correlation between $\Gamma$ and $\lambda_\mathrm{Edd}$ for the full sample, as well as for the super-Eddington subsample, with a slope of $\sim0.3$, consistent with that derived by, e.g, \citet{shemmer2008} and \citet{brightman2013}.
However, it is worth noting that for their study \citet{liu2021} did only consider spectra during high-flux states, in case of AGN with multiple X-ray observations, which may lead to a possible bias in estimating the slope of the best-fit relation for high-$\lambda_\mathrm{Edd}$ sources, given the trend of steeper-when-brighter typically observed in type-1 AGN \citep[e.g.][]{leighly1996, nandra1997, gibson2012, gliozzi2017, serafinelli2017}.

\begin{figure*}[t]
    \centering
    \includegraphics[width=2\columnwidth]{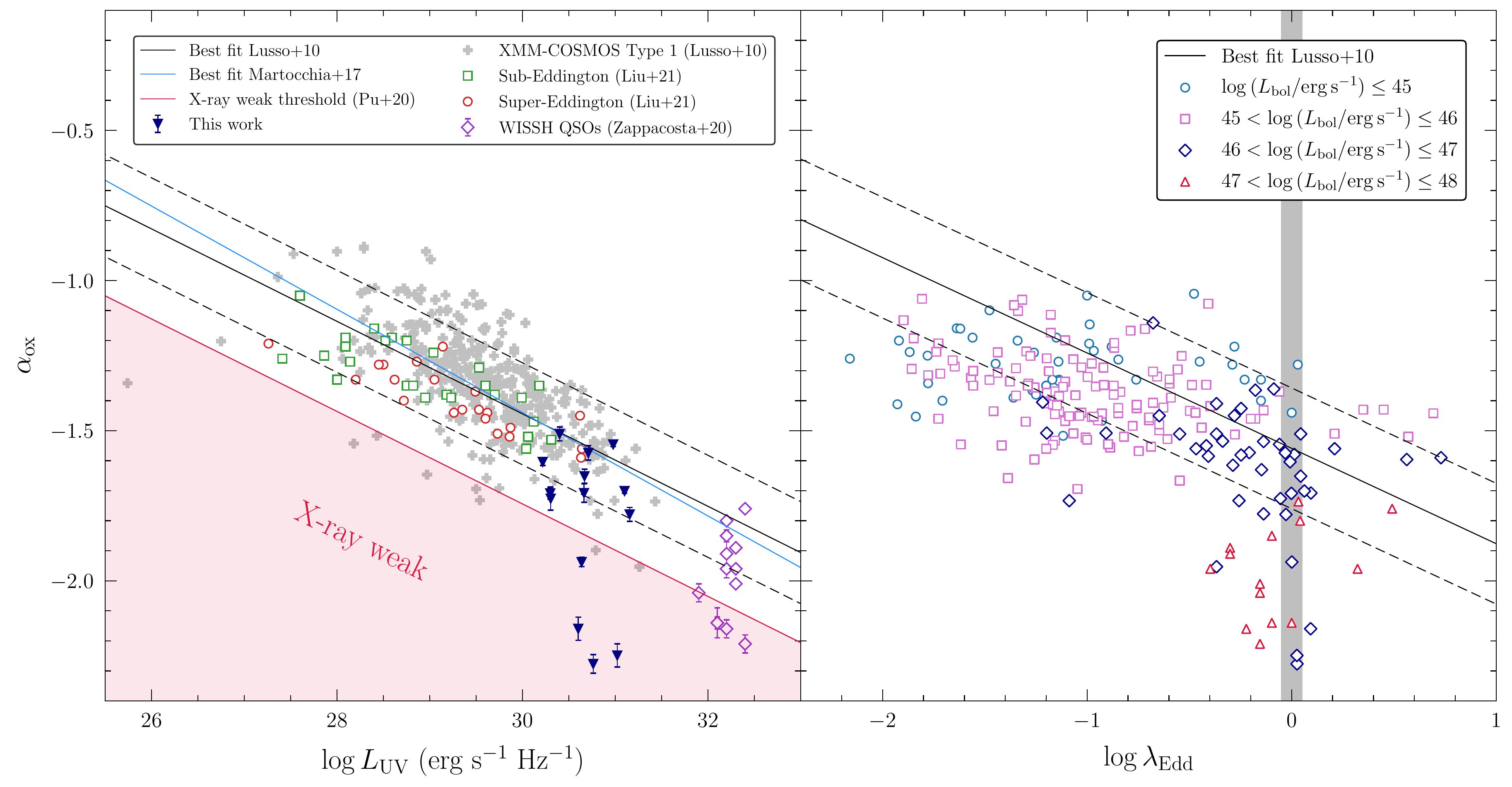}
    \caption{\emph{Left panel}: $\alpha_\mathrm{ox}$ vs $\log{L_\mathrm{UV}}$ for our high-$\lambda_\mathrm{Edd}$ AGN (blue triangles) compared to the XMM-COSMOS type-1 AGN (light grey crosses) of \citet{lusso2010}. Black solid line is the best-fit relation from the same authors. Black dashed lines describe the $1\sigma$ spread of the above relation. Blue solid line is the best-fit relation obtained from \citet{martocchia2017}. Green squares and red circles indicate the sub- and super-Eddington AGN from \citet{liu2021}, respectively. Purple diamonds are the 13 WISSH QSOs from \citet{zappacosta2020}. Red solid line marks the reference value for X-ray weakness, i.e. $\Delta\alpha_\mathrm{ox}\leq-0.3$  \citep{pu2020}. \emph{Right panel}: distribution of the same sources described in the adjacent panel, divided in two regimes of bolometric luminosity, in the $\alpha_\mathrm{ox} - \log{\lambda_\mathrm{Edd}}$ plane. For the XMM-COSMOS AGN sample, only those sources with available measurement of $\lambda_\mathrm{Edd}$  are included. Black solid line indicates the best-fit relation from \citet{lusso2010}. Black dashed lines describe the $1\sigma$ spread of the above relation. The shaded interval at $\log{\lambda_\mathrm{Edd}}\sim0$ is included to highlight that AGN with very similar values of \ledd\ can show a widespread distribution in $\alpha_\mathrm{ox}$ due to their different $L_\mathrm{bol}$.}
    \label{fig:aox_vs_UV}
\end{figure*}

\subsection{X-ray weakness}
\label{sub:weak}

Another relevant aspect emerging from the X-ray analysis of our high-\ledd\ AGN sample is shown in Fig. \ref{fig:kbolVSlbol}.
Our sources (blue triangles) are compared with the \kbol\ vs $\log{\lambda_\mathrm{Edd}}$ distribution of the XMM-COSMOS sample of type-1 AGN from \citet{lusso2010}, and 
the $k_\mathrm{bol,X} - \log{\lambda_\mathrm{Edd}}$ best-fit relation from \citet{duras2020}, which holds for both absorbed (type-2) and unabsorbed (type-1) AGN, is also reported.
The bulk of our sources is on average slightly above the range $\sim10-100$ expected for their $\log{\lambda_\mathrm{Edd}}$. Furthermore there are four high-\ledd\ AGN ($\sim30\%$ of the sample) exhibiting $k_\mathrm{bol,X}$ values significantly larger than expected and spanning a range between $\sim600-5000$, i.e.\ larger than the maximum $k_\mathrm{bol,X}$ of $\sim400$ predicted by the \citet{duras2020} relation for the most highly-accreting AGN.
These high-$k_\mathrm{bol,X}$ (hence X-ray weak) QSOs are J1103$+$41, J1127$+$64, J1245$+$33 and J0300–08, with the latter showing a $k_\mathrm{bol,X}$ of a few thousands.
The left panel of Fig. \ref{fig:aox_vs_UV} shows the distribution of the high-$\lambda_\mathrm{Edd}$ AGN in terms of the $\alpha_\mathrm{ox}$ parameter and the monochromatic optical/UV luminosity at 2500\,\AA, $\log{L_\mathrm{UV}}$, and provides an additional piece of evidence to support this scenario.
The same four QSOs are clearly offset from the $\alpha_\mathrm{ox}$ vs $\log{L_\mathrm{UV}}$ relation found for large AGN samples \citep[e.g.][]{just2007, lusso2010, vagnetti2010, martocchia2017}, falling in the region of the plane defined by $\Delta\alpha_\mathrm{ox}\leq-0.3$, where $\Delta\alpha_\mathrm{ox}$ represents the difference between the observed value of $\alpha_\mathrm{ox}$ and the expected value from the \citet{lusso2010} relation. 
In their recent work, \citet{pu2020} argued that those sources with $\Delta\alpha_\mathrm{ox}\leq-0.3$ can be reasonably classified as X-ray weak AGN \citep[e.g.][]{brandt2000,miniutti2009}, given that such a $\Delta\alpha_\mathrm{ox}$ implies a $L_{2-10\,\mathrm{keV}}$ weaker by a factor of $f_{\rm weak} = 10^{-\Delta\alpha_\mathrm{ox}/0.3838} \geq 6$ than the bulk of the AGN population at a similar $\log{L_\mathrm{UV}}$. Specifically, the X-ray weakness factors $f_{\rm weak}$ measured for J1103$+$41, J1127$+$64, J1245$+$33 and J0300–08 span from $\sim 11$ to $\sim 75$, with $\Delta\alpha_\mathrm{ox}$ comprised between $-0.7$ and $-0.4$.
It is worth noting that the fraction of QSOs with $f_{\rm weak} \geq 6$ in our sample is 29$^{+22}_{-14}$\% (with errors estimated according to \citealt{gehrels1986}), which is much larger (and inconsistent at the $\sim99\%$ confidence level) than that derived by \citet{pu2020} for a sample of 1825 type-1 QSOs from SDSS, i.e. $5.8\pm0.7\%$.
The $\alpha_\mathrm{ox}-\log\lambda_\mathrm{Edd}$ distribution of our fourteen high-$\lambda_\mathrm{Edd}$ AGN, the XMM-COSMOS type-1 AGN of \citet{lusso2010}, the sub- and super-Eddington AGN of \citet{liu2021} and the thirteen hyper-luminous WISSH QSOs of \citet{zappacosta2020}, divided in different regimes of bolometric luminosity, is shown in the right panel of Fig.\ \ref{fig:aox_vs_UV}.
A large scatter around the best-fit relation from \citet{lusso2010} is clearly visible in Fig.\ \ref{fig:aox_vs_UV}, and is probably reflecting the difficulty to derive a tight $\alpha_\mathrm{ox}-\log\lambda_\mathrm{Edd}$ relation for large samples of sources with different bolometric luminosities.  

\begin{figure}[t]
\includegraphics[width=\columnwidth]{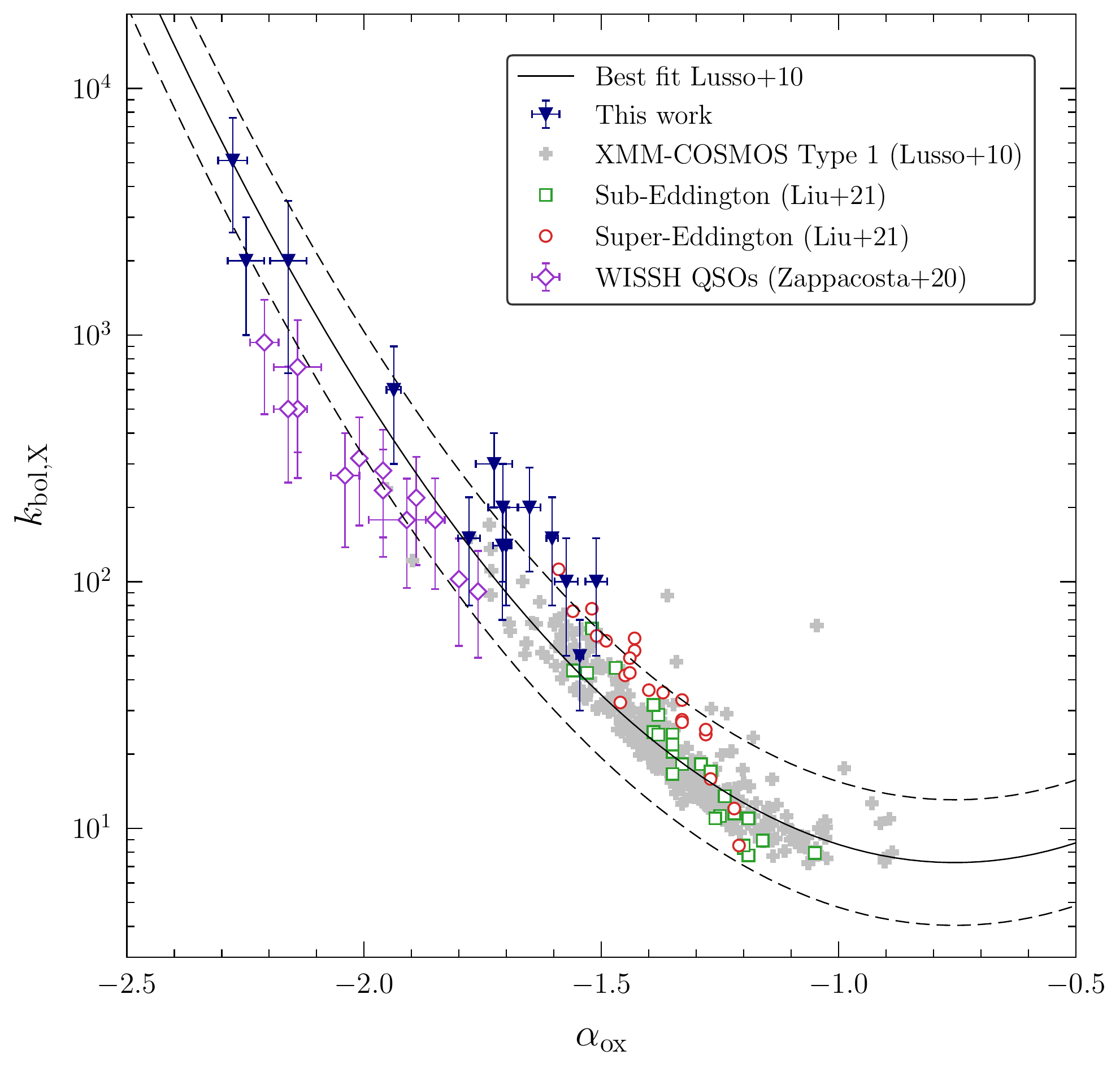}
\caption{$k_\mathrm{bol,X}$ vs $\alpha_\mathrm{ox}$ for the high-$\lambda_\mathrm{Edd}$ AGN in our sample (blue triangles) compared to the XMM-COSMOS type-1 AGN (light grey crosses) of \citet{lusso2010}. Black solid line refers to their best-fit relation, with the black dashed lines describing its $1\sigma$ spread. Green squares and red circles indicate the sub- and super-Eddington AGN from \citet{liu2021}, respectively. Purple diamonds are the hyper-luminous WISSH QSOs from \citet{zappacosta2020}.}
\label{fig:aox_vs_kx}
\end{figure}

Finally, Fig.\ \ref{fig:aox_vs_kx} shows the distribution of the fourteen highly-accreting QSOs in the plane $k_\mathrm{bol,X} - \alpha_\mathrm{ox}$, along with the relation derived for the COSMOS type-1 AGN that exhibit $k_\mathrm{bol,X}$ in the range $\approx$ $10-100$. Typical local AGN in the sample of \citet{liu2021} are also presented for comparison.
The four X-ray weak high-\ledd\ AGN are also located near the $k_\mathrm{bol,X} - \alpha_\mathrm{ox}$ relation in \citet{lusso2010}, being these two quantities closely related with each other, both of them indicating the relative strength between the X-ray emission produced in the corona and the UV emission from the accretion disc.
Furthermore, it appears that the addition of a large number of X-ray weak sources might be useful to establish the exact $k_\mathrm{bol,X} - \alpha_\mathrm{ox}$ relation in a broad range of $\alpha_\mathrm{ox}$, with the latter serving as a proxy for \lbol. 

\section{Discussion}
\label{sec:discussion}

\subsection{Properties of the X-ray continuum and X-ray weakness}
\label{sub:5.1}

The 14 high-\ledd\ AGN analysed in this paper are drawn from the larger MS14 sample of intermediate-redshift, highly-accreting quasars, which exhibit remarkably homogeneous optical properties (e.g. extremely strong $\mathrm{Fe{\,\scriptstyle II}}$ emission, weak [$\mathrm{O{\,\scriptstyle III}}$] emission, FWHM(H$\beta$) $<$ 4000 km s$^{-1}$). Furthermore, our sample has a narrow distribution in both redshift ($\sigma_z$ $\sim$ 0.1) centred at $z \sim$  0.57, and Eddington ratio ($\sigma_{\lambda_\mathrm{Edd}}$ $\sim$ 0.1) centred at  $\lambda_\mathrm{Edd} \sim 1.1$, while luminosities and SMBH masses are clustered around $\log{(L_\mathrm{bol}/\mathrm{erg\,s}^{-1})} \approx 46.5$ and $\log{(M_\mathrm{BH}/M_\odot)}$ $\approx$ 8.3, respectively (see Tab.\ \ref{tab:AGNample}).
The main result emerging from our analysis is that, despite the homogeneous optical and SMBH accretion properties, the X-ray properties of these xA quasars
appear to be quite heterogeneous. This, in turn, indicates that the structure and efficiency of the X-ray corona and innermost accretion flow in these AGN  cannot be considered a uniform and distinctive feature for reaching a maximised radiative output per unit mass. In this sense, the scatter of the primary continuum slope measured for our highly-accreting  quasars as a function of the logarithm of $\lambda_\mathrm{Edd}$ (Fig.\ \ref{fig:gammaVSedd}) suggests caution in the use of $\Gamma$ for providing a reliable estimate of the SMBH mass based on the X-ray spectrum.
Indeed, at these high-Eddington ratios this may be telling us something about the structure
of the disc/corona complex, possibly indicating a (time-variable?) redistribution of disc accretion power
dissipated in the corona, such as a variation in the inner truncation radius of the accretion disc
or the geometry of the disc/corona complex \citep[e.g.][]{kubota2018}.  This, in turn, would affect the slope
of the hard X-ray emission. In particular, a larger number of sources with $M_\mathrm{BH}$ at the high-end of the mass distribution (i.e. $\geq$ 10$^{9}$ $M_\odot$) and a well-constrained hard X-ray photon index is needed to evaluate the presence of possible biases due to a dependence of $\Gamma$ on $M_\mathrm{BH}$ 
\citep[e.g.][]{martocchia2017}.

Another piece of evidence supporting a large variety of the X-ray properties of our high-$\lambda_\mathrm{Edd}$ quasars comes from the wide distributions of $k_\mathrm{bol,X} - \log{\lambda_\mathrm{Edd}}$ and $\alpha_\mathrm{ox}-\log{L_\mathrm{UV}}$ reported in Fig.\ \ref{fig:kbolVSlbol} and \ref{fig:aox_vs_UV}. In particular, four sources exhibit large values of $k_\mathrm{bol,X} $ and small $\Delta\alpha_\mathrm{ox}$, and they can be classified as X-ray weak quasars. 
For these sources, we measure an X-ray emission about a factor of $\sim 10-80$ fainter than the bulk of the AGN population at similar UV luminosities. 
Bearing in mind the uncertainties due to the limited number of sources, the fraction of QSOs with $f_{\rm weak}\geq6$ in our sample is $29^{+22}_{-14}\,\%$, which is much larger than that derived for the AGN general population \citep[e.g.][]{pu2020}, suggesting a possible change in the properties of the accretion disc-corona system in some objects radiating close (or above) to the Eddington limit, possibly related to the increasing importance of radiation pressure in shaping the efficiency of the accretion process \citep[e.g.][]{proga2004, sadowski2016, zappacosta2020, nomura2020}.
We note that X-ray weakness in highly-accreting quasars is often attributed to absorption. This has typically been the case for BAL quasars (\citealt{gallagher2002}; \citealt{piconcelli2005}; but see also \citealt{morabito2014} and \citealt{teng2014}) and weak emission-line quasars (WLQs; e.g.\ \citealt{luo2015}; \citealt{ni2018}), for which the presence of a nuclear shielding gas component with a large covering factor has been proposed to explain the weak broad emission lines  and wind acceleration
without over-ionisation \citep[e.g.][]{wu2011}.
In the case of high-\ledd\ AGN, this shielding gas may be identified with an optically and geometrically thick accretion disc \citep{luo2015}, or with a failed wind resulting from stalling of line-driven disc winds \citep[e.g.][]{proga2004, proga2005, nomura2020}.  In any case, 
the shielding gas should also be responsible for the X-ray absorption along our line of sight to the nucleus.
Both BAL quasars and WLQs are indeed characterised by hard X-ray spectra, when the broadband continuum is modelled with a single power law. Indeed, on average, the resulting effective photon index is usually $\langle\Gamma_\mathrm{eff}\rangle \sim 1.2$. \citet{pu2020}, for instance, indicated $\Gamma_\mathrm{eff}=1.26$ as the threshold below which a quasar is potentially obscured by cold absorption.
However,  by modelling the $0.3-10$ keV X-ray continuum with a power law, we find none of our four high-\ledd\, X-ray weak sources to have flat photon indices compatible with those typical of  WLQs and other absorbed quasars. Indeed, they show very steep $\Gamma_\mathrm{eff}\sim 2.2-3$, suggesting that intervening cold obscuration is not the likely cause of their X-ray weakness, which may therefore be an intrinsic property of these AGN. 
Nonetheless, we cannot rule out the presence of a highly-ionised gas cloud with a large $N_\mathrm{H}$ and small covering factor along our line of sight which may scatter off a large amount of photons, causing a decrease of the X-ray flux without altering the spectral shape. However, this \emph{ad hoc} scenario is difficult to test directly and seems quite unlikely to apply to all the four X-ray weak QSOs in our sample.
In addition, \citet{miniutti2012} proposed a couple of alternative scenarios for the 
intrinsic X-ray weakness of the $z\sim 0.4$, high-\ledd\ quasar PHL 1092, which is one of the most extreme X-ray weak quasars with no BAL and an almost constant flux in the UV. In the former scenario, the size of the X-ray corona changes as a function of the X-ray flux, and a very compact X-ray corona further shrinks down to the innermost stable circular orbit around the SMBH during X-ray weak states. In the second scenario, the corona is located within few gravitational radii from the SMBH at all flux levels, and light-bending effects give rise to a disc reflection-dominated X-ray spectrum in extreme low-flux states. However, this hypothesis can be tested only by analysing additional, deeper X-ray observations (preferably including data at $>10$ keV) to reveal typical relativistic reflection spectral features. Similary, light-bending effects have also been invoked by \citet{ni2020} as a viable explanation to justify the extreme X-ray variability of the $z=1.935$, WLQ SDSS J153913.47+395423.4. This source was found to be in an X-ray weak state, before its flux raised by a factor of $\sim 20$, approximately six years after its first observation (corresponding to $\sim2$ years in the QSO rest frame). Alternatively, the authors also suggested that the X-ray weakness could be caused by nuclear shielding, probably due to the inner thick disc, which may intercept the line of sight to the central source.

High-\ledd\ AGN are also likely to accelerate powerful outflows, which can provide an additional effect contributing to weaken the X-ray emission.  \citet{laurenti2021} analysed the UFO in the NLSy1 galaxy PG 1448+273 with $\lambda_\mathrm{Edd}$ $\sim$ 0.75. They found that this source underwent large variations of $\alpha_\mathrm{ox}$, with a maximum offset of $\Delta\alpha_\mathrm{ox}=-0.7$, after the UFO was detected. As a result, PG 1448+263 was characterised by a remarkable X-ray weakness for some months, before the flux finally returned to the same level as prior to the UFO detection. In this case, the authors suggested that such a powerful disc wind could be responsible of the observed X-ray weakness, by removing a large amount of the infalling material in the innermost part of the accretion disc and, thus, reducing the flux of seed photons towards the corona.

In addition, \citet{zappacosta2020} recently reported a relation between $L_{2-10}$ and the velocity shift $v_{\mathrm{C{\,\scriptscriptstyle IV}}}$ of the $\mathrm{C{\,\scriptstyle IV}}$ emission line profile over $\sim$ 1.5 dex in $L_{2-10\,\mathrm{keV}}$ for a sample of thirteen very luminous quasars at $z >$ 2, with $\lambda_\mathrm{Edd}\sim0.5-3$ estimated from H$\beta$-based $M_\mathrm{BH}$. These $\mathrm{C{\,\scriptstyle IV}}$ shifts are interpreted in terms  of winds produced at accretion-disc scale, and the fastest winds appear to be associated with the lowest hard X-ray luminosities and steepest $\alpha_\mathrm{ox}$ values. 
Interestingly, \citet{proga2005} suggests that in case of highly-accreting AGN,  failed winds, which are inevitably produced along with UV line-driven winds, may affect the density of the X-ray corona and weaken its inverse Compton X-ray emission.

\begin{figure}[t]
  \centering
  \includegraphics[width=\columnwidth]{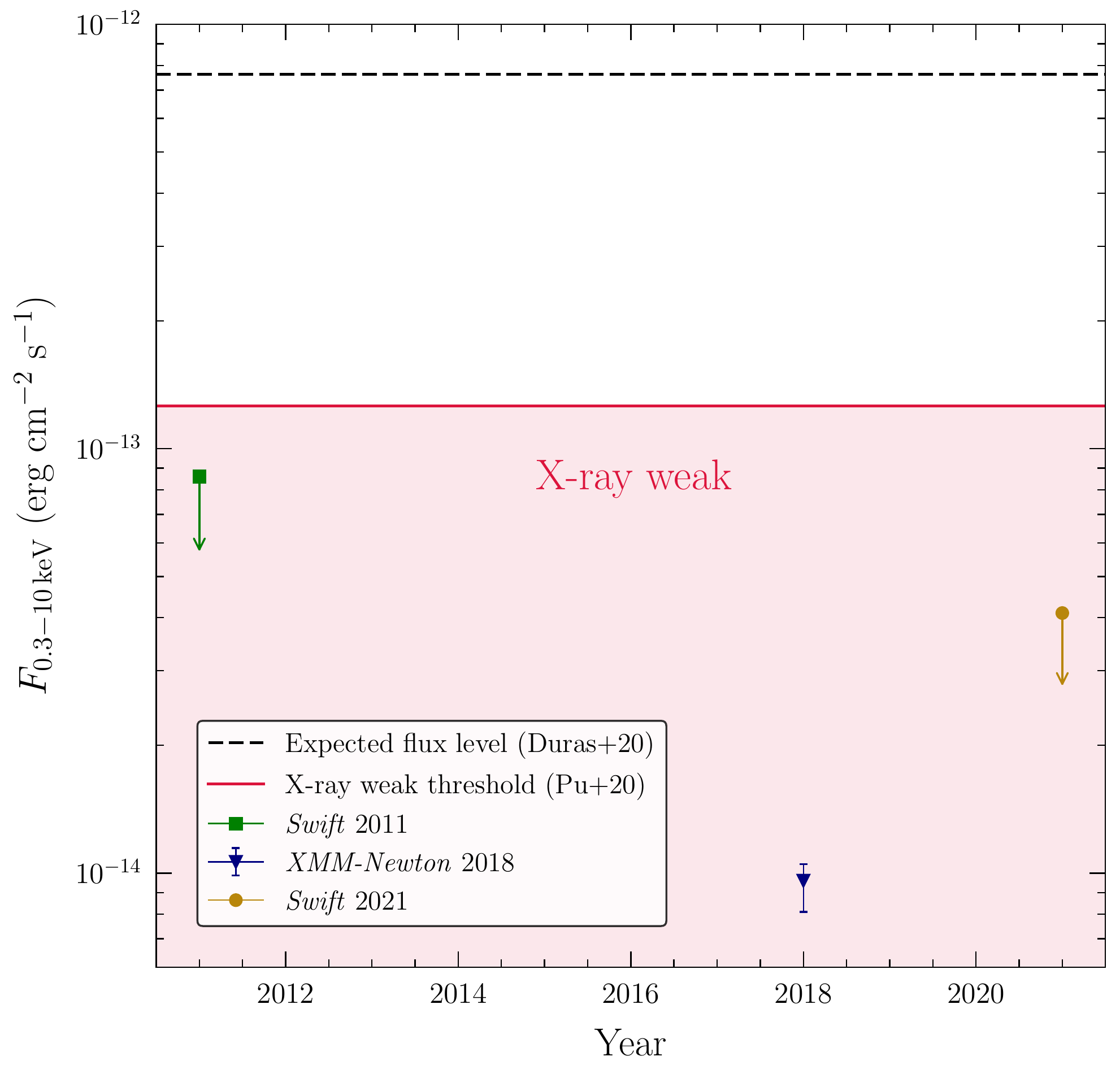}
  \caption{Light curve of J0300$-$08 in terms of the broadband $E=0.3-10$ keV flux. Black dashed line represents the flux level of $F_{0.3-10\,\mathrm{keV}} = 7.6\times10^{-13}$ erg cm$^{-2}$ s$^{-1}$ expected from its bolometric luminosity, once the $k_\mathrm{bol,X}-\log{L_\mathrm{bol}}$ relation from \citet{duras2020} is assumed. Red solid line indicates the flux reference value for X-ray weakness corresponding to $\Delta\alpha_\mathrm{ox}\leq-0.3$ \citep{pu2020}. The $3\sigma$ upper limits from the \emph{Swift} observations of 2011 and 2021 are shown in green and gold, respectively. The flux  from the 2018 \emph{XMM-Newton} observation analysed in this work is marked in blue.}
  \label{fig:fluxes_J0300-08}
\end{figure}

\citet{zappacosta2020} found that X-ray weak sources (i.e. those with $\Delta\alpha_\mathrm{ox} \leq -0.3$) represent a fraction of $\sim40\%$ of their sample and all exhibit $v_{\mathrm{C{\,\scriptscriptstyle IV}}} \geq 5000$ km s$^{-1}$.
\citet{nardini2019} have recently analysed the XMM spectra of a sample of 30 intrinsically blue, non-BAL, $L_\mathrm{bol} \gtrsim 10^{47}$ erg s$^{-1}$ quasars at $z \sim 3-3.3$ which are likely to shine at $\lambda_\mathrm{Edd} \sim 1$ \citep[see also][]{lusso2021}. They reported a fraction of $\sim25\%$ of X-ray weak sources, according to a $-0.4<\Delta\alpha_\mathrm{ox} < -0.2$.
Interestingly, our finding of a similar fraction of X-ray weak sources in our sample of $z\sim0.6$, $L_\mathrm{bol}\sim10^{46}$ erg s$^{-1}$, optically-selected quasars therefore suggests that (i) a high frequency of intrinsically X-ray weak sources is not a distinctive feature of the population of ultra-luminous,
high-$z$ quasars and (ii) a high-\ledd\ ratio may imply favourable accretion disc conditions for enhancing the
probability of catching an AGN in an X-ray weak state.

Finally, we were recently awarded time on the \emph{Neil Gehrels Swift} Observatory (\citealt{gehrels2004}, hereinafter \emph{Swift}) for monitoring the X-ray emission of the four X-ray weak sources discovered by the XMM-Newton observations presented here and tracking down possible rises of their X-ray emission (PI: E.\ Piconcelli).
The first observed source is J0300$-$08, which was targeted by \emph{Swift} for 16 ks on March 2021. The source was not detected, with a 3$\sigma$ upper-limit on the $0.3-10$ keV flux of 4.1 $\times10^{-14}$ erg cm$^{-2}$ s$^{-1}$, indicating that this quasar still exhibits a weak X-ray emission.
Fig.\ \ref{fig:fluxes_J0300-08} shows this value along with the fluxes derived from the 2018 \emph{XMM-Newton} observation and from an additional \emph{Swift} 6 ks archival observation performed in 2011, for which we derive an upper-limit of  8.6 $\times10^{-14}$ erg cm$^{-2}$ s$^{-1}$.
According to the $k_\mathrm{bol,X}-\log{L_\mathrm{bol}}$ best-fit relation from \citet{duras2020}, from the $\log{L_\mathrm{bol}}$ listed in Tab.\ \ref{tab:AGNample} for J0300$-$08 we would expect a flux of around $F_{0.3-10\,\mathrm{keV}} \sim 8\times10^{-13}$ erg cm$^{-2}$ s$^{-1}$.
All the measured X-ray fluxes are well below the level expected for the bulk of type-1 AGN with comparable $\log{L_\mathrm{bol}}$, as shown in Fig.\ \ref{fig:fluxes_J0300-08}. This may indicate that J0300$-$08 has likely been in a predominantly low-flux state during the past decade, suggesting that  the X-ray weakness regime may persist over a timescale of several years.

\subsection{Relations involving $\lambda_\mathrm{Edd}$}

The results presented in Sect.\ \ref{sec:results} highlight the importance of enlarging the number of high-\ledd\ quasars with dedicated X-ray observations. This would enable
new opportunities to investigate the high-\ledd\ phenomenon at large and extend the dynamical range for any possible correlation involving X-ray derived parameters and the Eddington ratio.
Specifically, Fig.\ \ref{fig:kbolVSlbol} shows that the relation between $k_\mathrm{bol,X} - \log{\lambda_\mathrm{Edd}}$ by \citet{duras2020} seems to provide a fair description of the correlation for the the bulk of the high-\ledd\ AGN population, i.e. for both small- and large-$M_\mathrm{BH}$ sources, except for the X-ray weak sources that deviate considerably from the best-fit relation. Conversely, the large scatter in the right panel of Fig.\ \ref{fig:aox_vs_UV} indicates that only additional investigations involving a large number of high-\ledd\ AGN at different luminosity regimes may deepen our understanding of the relation between $\alpha_\mathrm{ox}$ and $\log{\lambda_\mathrm{Edd}}$. In particular, as shown in the same figure, sources with very similar values of $\log{\lambda_\mathrm{Edd}}$ can exhibit a widespread distribution of $\alpha_\mathrm{ox}$ values according to their $L_\mathrm{bol}$ ($\approx$ $L_\mathrm{UV}$ in case of type-1 AGN). This suggests that the presence of a large scatter observed in the $\alpha_\mathrm{ox}-\log{\lambda_\mathrm{Edd}}$ plane is intrinsic, and the accurate description of this relation for the whole AGN population should take into account this dependence on $L_\mathrm{bol}$. This may also explain the opposite conclusions on the existence of a significant correlation between $\alpha_\mathrm{ox}-\log{\lambda_\mathrm{Edd}}$ reported by different studies so far \citep[e.g.][]{vasudevan2007, shemmer2008, lusso2010, chiaraluce2018}.

\begin{figure}[t]
  \centering
  \includegraphics[width=\columnwidth]{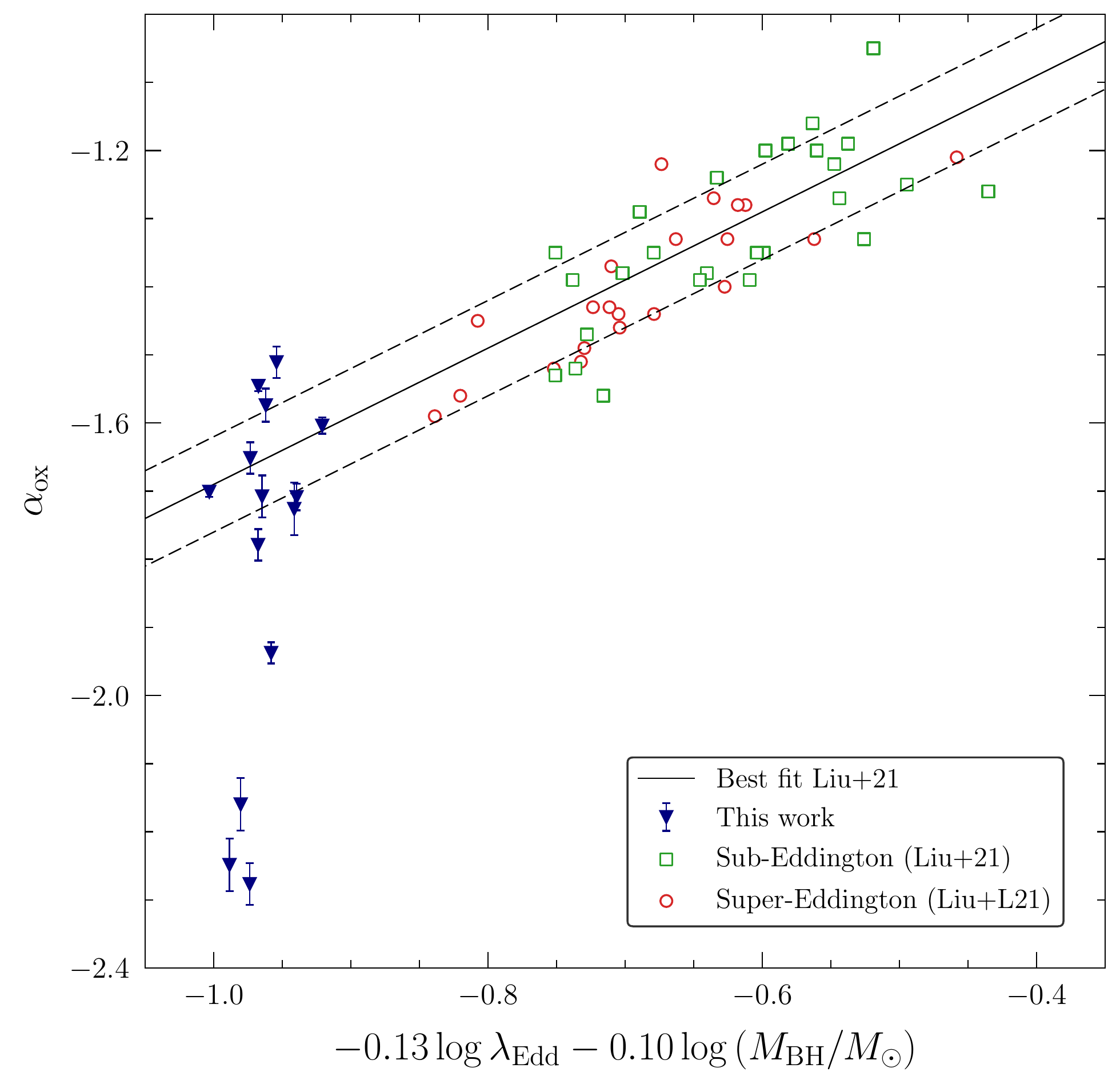}
  \caption{$\alpha_\mathrm{ox}-\lambda_\mathrm{Edd}- M_\mathrm{BH}$ plane introduced by \citet{liu2021}. Green squares and red circles indicate their sub- and super-Eddington AGN, respectively. Black solid and dashed lines describe the best-fit relation and its spread as indicated by the same authors, respectively. We extend their relation towards higher accretion rates and black hole masses by including our high-\ledd\ AGN (blue triangles). }
  \label{fig:fundam_plane}
\end{figure}

Furthermore, \citet{liu2021} have recently proposed the possible existence of a more complex scenario for the relation between $\alpha_\mathrm{ox}$ and $\log{\lambda_\mathrm{Edd}}$, which also involves another fundamental parameter, i.e. $M_\mathrm{BH}$.
On the one hand they found that both sub- and super-Eddington AGN follow the same $\alpha_\mathrm{ox} - \log{L_\mathrm{UV}}$ relation, which may indicate that the properties of the accretion disc-corona system do not show large differences as a function of the accretion rate for the sources in their sample, which are all observed in a high-flux state.
On the other hand, \citet{liu2021} found that the anti-correlation between $\alpha_\mathrm{ox}$ and $\log{\lambda_\mathrm{Edd}}$ is more significant when the two subsamples of sub- and super-Eddington AGN are considered separately instead of the entire AGN sample. A similar behaviour is also observed for $\alpha_\mathrm{ox}$ as a function of $\log{(M_\mathrm{BH}/M_\odot)}$.
Interestingly, by means of a partial correlation analysis, \citet{liu2021}  obtained a statistically significant non-linear relation between $\alpha_\mathrm{ox}$ and both $\lambda_\mathrm{Edd}$ and $M_\mathrm{BH}$, in the form $\alpha_\mathrm{ox} = \beta\,\log{\lambda_\mathrm{Edd}} + \gamma\,\log{(M_\mathrm{BH}/M_\odot)} + \delta$, with $\beta\simeq-0.13$, $\gamma\simeq-0.10$ and $\delta\simeq-0.69$.
This best fit is shown in Fig.\ \ref{fig:fundam_plane} along with the data from the \citet{liu2021} sub- and super-Eddington samples.
The relation appears to hold once extended towards steeper $\alpha_\mathrm{ox}$ and larger $M_\mathrm{BH}$ by including our high-\ledd\ quasars.
Clearly, being based on sources observed in a high-flux state, the \citet{liu2021} relation cannot describe the behaviour of the four X-ray weak sources in our sample. 
On the basis of their data \citet{liu2021} were not able to determine whether this  $\alpha_\mathrm{ox}-\lambda_\mathrm{Edd}- M_\mathrm{BH}$ relation represents the fundamental plane, or a secondary effect of the  $\alpha_\mathrm{ox} -\log{L_\mathrm{UV}}$ relation. Moreover, $\alpha_\mathrm{ox}$, $\lambda_\mathrm{Edd}$, and $M_\mathrm{BH}$ are all intimately connected and cannot be considered as independent variables. This suggests to use caution when determining relations that involve these quantities. However, the investigation on the nature of the physical driver of these relations will certainly be a hotly-debated topic in the future.

Finally, it is worth noting that although most of our high-$\lambda_\mathrm{Edd}$ QSOs are basically unobscured in the optical, five sources are characterised by an internal reddening E($B-V$) $\geq 0.05$, and two of them having E($B-V$) $\geq0.1$ (see Tab. \ref{tab:optical_data}).
This is particularly intriguing since \citet{krawczyk2015} analysed both the reddening and extinction of $\sim35000$ type-1 QSOs from the SDSS, with $z\leq5$ including both BAL and non-BAL quasars. They concluded that only $9\%$ ($35\%$) of the non-BAL (BAL) AGN has an E($B-V$) $\geq 0.05$, and $3\%$ ($11\%$) has an E($B-V$) $\geq0.1$.
This means that, independently of their classification as non-BAL or BAL QSOs, the five reddened, high-\ledd\ AGN in our sample belong to the right-hand tail of the reddening distribution of type-1 AGN. 
For what concerns the X-ray appearance  of these five sources, we do not find any trend with E($B-V$), as their main X-ray spectral properties are widely heterogeneous. Indeed, they show a power-law continuum with photon indices comprised between $1.3 \leq \Gamma \leq 2.3$, values of $\alpha_\mathrm{ox}$ between $\sim -1.6$ and $\sim -1.9$, and the $R_\mathrm{S/P}$ parameter ranges from $\sim0.5$ to $\sim5.5$. In addition, only one out of five reddened sources (namely J1103+41) can be classified as X-ray weak quasar, showing a $k_\mathrm{bol,X} = 600\pm300$.

\section{Summary and conclusion}
\label{sec:conclusion}

In this study, we have investigated the X-ray and optical/UV properties of 14 high-$\lambda_\mathrm{Edd}$, radio-quiet quasars observed by \emph{XMM-Newton} (see Sect.\ \ref{sec:data}).
These sources belong to the sample of intermediate-redshift, highly-accreting quasars presented in MS14, and are characterised by homogeneous spectral optical and SMBH (i.e. $M_\mathrm{BH}$ and $\lambda_\mathrm{Edd}$) properties. The main results can be summarised as follows:

\begin{enumerate}
    \item Our high-\ledd\ quasars exhibit significant differences in their  X-ray properties.  Specifically, we find a large scatter in the distribution of the continuum slope, with a non-negligible fraction of these AGN showing $\Gamma\leq1.6$ (see Fig.\ \ref{fig:gammaVSedd}).
Bearing in mind the limited number of sources analysed here, this result suggests that the X-ray properties of high-\ledd\ AGN can be more heterogeneous than previously reported.
In this sense, the observed offsets from the  $\Gamma$ vs \ledd\ relations found in the past \citep[e.g.][]{shemmer2008, brightman2013,huang2020} can be a manifestation of distinctive physical properties of the inner accretion disc--X-ray corona in these AGN.
    
    \item We also find that $\sim30\%$ of the  high-$\lambda_\mathrm{Edd}$ sources considered in our study show an  X-ray weakness factor $f_{\rm weak} > 10$ corresponding to $\Delta\alpha_\mathrm{ox}$ comprised between $-0.7$ and $-0.4$ (see Figs.\ \ref{fig:aox_vs_UV} and \ref{fig:aox_vs_kx}).
They can be therefore classified as X-ray weak quasars. The steep broad-band  X-ray spectral shape (i.e. $\Gamma_{\rm eff} \geq 2.2$; see Sect.\ \ref{sub:5.1}) seems to rule out absorption as the cause of their reduced X-ray emission, lending support to an intrinsic X-ray weakness.
Interestingly, this fraction is similar to that reported by \citet{nardini2019} and \citet{zappacosta2020} who studied samples of very luminous ($L_\mathrm{bol}\gtrsim10^{47}$ erg s$^{-1}$) quasars at $z \sim2-3$, which are likely to shine at $\lambda_\mathrm{Edd}\gtrsim1$. This indicates that a high \ledd\ might be a key parameter for triggering a weak X-ray corona state. A systematic X-ray/UV study of large samples of $\lambda_\mathrm{Edd}\geq1$ AGN would therefore be useful to shed light on the distinctive nuclear properties of sources undergoing an intrinsically X-ray weak phase.
Furthermore, a recent \emph{Swift} monitoring of J0300$-$08, one of the four X-ray weak QSOs in our sample, reveals that the source was still in a low-flux state in March 2021. Combining this measurement with the 2018 \emph{XMM-Newton} and an archival 2011 \emph{Swift} observation, it appears that J0300$-$08 has likely been going through a period of intrinsic X-ray weakness for nearly a decade (see Fig.\ \ref{fig:fluxes_J0300-08}).
    
\item  The analysis of the optical spectra of the fourteen high-\ledd\ quasars reveals that five sources have a reddening E($B-V$) $\geq 0.05$. Such a fraction of reddened sources is higher than that typically measured for the non-BAL quasar population, and we do not report any apparent link with specific X-ray spectral and $\Delta\alpha_\mathrm{ox}$ properties. Spectroscopy in the UV band would be very useful to detect the possible presence of BAL features in these reddened, highly-accreting quasars at intermediate redshifts and shed light on their nature.

\end{enumerate}

\noindent We note that, given the relatively bright $0.5-2$ keV fluxes of the quasars in the present sample and the expected soft X-ray flux limits of the eROSITA surveys (eRASS:1, i.e. first full-sky data release: $\approx4.5\times10^{-14}$~erg~cm$^{-2}$~s$^{-1}$; eRASS:8, final full-sky data release: $\approx1\times10^{-14}$~erg~cm$^{-2}$~s$^{-1}$; \citealt{merloni2012}), we can expect that $\approx70-80$\% of our highly-accreting quasars will be detected by eROSITA. 
However, the paucity of X-ray photons provided by eROSITA in survey mode will strongly limit any detailed X-ray spectral analysis, thus calling for more sensitive pointed X-ray observations.

Finally, the results presented in this paper inspire the following lines of future investigation: (i) a monitoring of the X-ray flux of the high-\ledd\ quasars analysed here, both X-ray ``normal'' and ``weak'' ones, over multiple timescales (i.e. months to years). This would allow us to get useful insights on the frequency and duration of the transitions between these two states; (ii) deep X-ray spectroscopy of the X-ray weak quasars is also desirable to provide unambiguous constraints on the spectral shape and the possible presence of some absorption along our line of sight to the nucleus.

\begin{acknowledgements}
We thank the anonymous referee for her/his useful comments. This work is based on observations obtained with \emph{XMM-Newton}, an ESA science mission with instruments and contributions directly funded by ESA Member States and NASA. Part of this work is based on archival data, software and online services provided by the Space Science Data Center - ASI. This work has been partially supported by the ASI-INAF program I/004/11/4.
ML acknowledges financial support from the Ph.D. programme in Astronomy, Astrophysics and Space Science supported by MIUR (Ministero dell'Istruzione, dell'Universit\`a e della Ricerca). EP, LZ, FT, SB, AL and GV acknowledge financial support under ASI/INAF contract 2017-14-H.0.
EP, SB, MB and GV acknowledge support from PRIN MIUR project "Black Hole winds and the Baryon Life Cycle of Galaxies: the stone-guest at the galaxy evolution supper", contract \#2017PH3WAT. AdO acknowledges financial support from the State Agency for Research of the Spanish MCIU through the project PID2019-106027GB-C41 and the ``Center of Excellence Severo Ochoa'' award to the Instituto de Astrof\'isica de Andaluc\'ia (SEV-2017-0709). RM acknowledges the financial support of INAF (Istituto Nazionale di Astrofisica), Osservatorio Astronomico di Roma, ASI (Agenzia Spaziale Italiana) under contract to INAF: ASI 2014-049-R.0 dedicated to SSDC. CR acknowledges support from the Fondecyt Iniciacion grant 11190831.
\end{acknowledgements}

\bibliographystyle{aa} 
\bibliography{hiedd.bib}

\appendix

\onecolumn\section{X-ray spectra of high-\ledd\ AGN}\label{sec:appA}

Fig.\ \ref{fig:spectra} shows the results of the broadband spectral fits to the EPIC spectrum  (pn, MOS1 and MOS2) of the 14 high-\ledd\ quasars listed in Tab.\ \ref{tab:AGNample}. The best fit to the bulk of the spectra is provided by a model consisting of a power law and a blackbody component, both modified for Galactic absorption (\texttt{tbabs $\cdot$ (zpowerlw + zbbody)} in the XSPEC notation).

\noindent The only exceptions are J0809+46 and J1048+31.
The former shows no excess emission in the soft X-rays where the spectrum, instead, is dominated by an ionised absorption component with high column density (Piconcelli et al., \emph{in prep.}), while the latter requires an additional component  (i.e.\ a Gaussian line) to account for the Fe K$\alpha$ emission line. Further details on the X-ray spectral analysis can be found in Sect.\ \ref{sec:spec}.

\vspace{1cm}

\begin{figure*}[!htbp]
  \centering
  \subfigure{\includegraphics[width=0.4\textwidth]{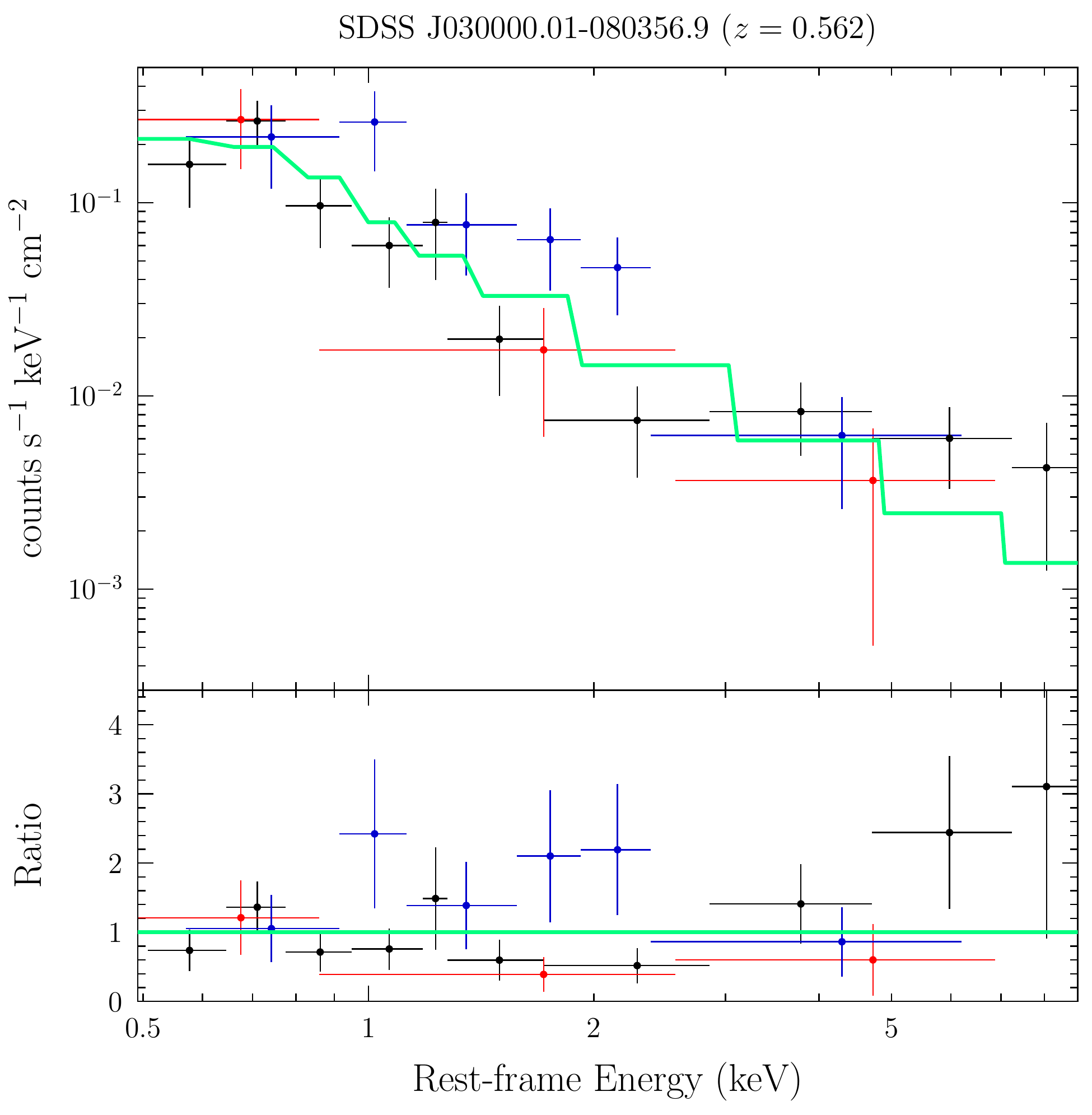}} \hfill
  \subfigure{\includegraphics[width=0.4\textwidth]{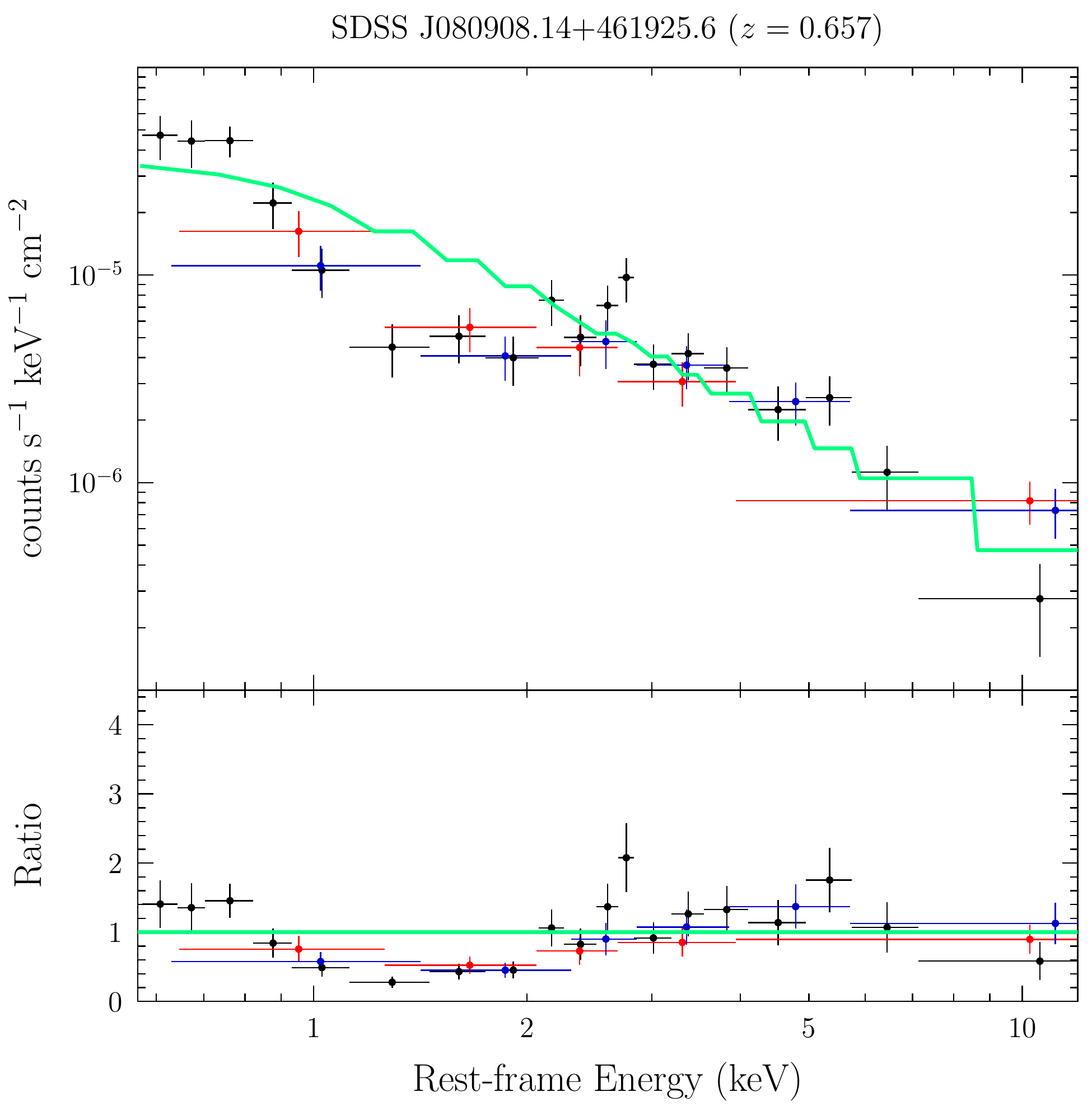}} \hfill\null
  
  \subfigure{\includegraphics[width=0.4\textwidth]{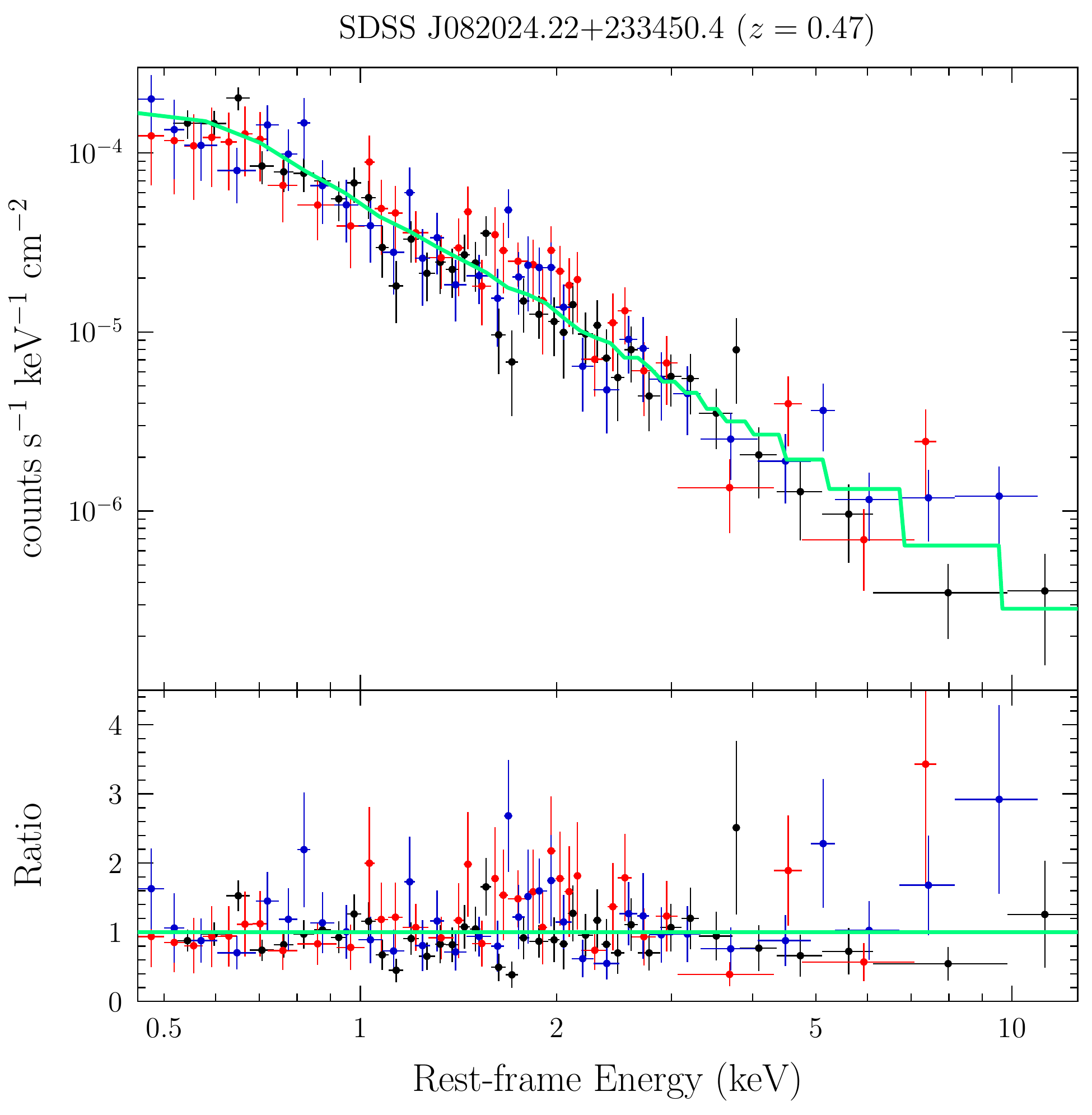}} \hfill
  \subfigure{\includegraphics[width=0.4\textwidth]{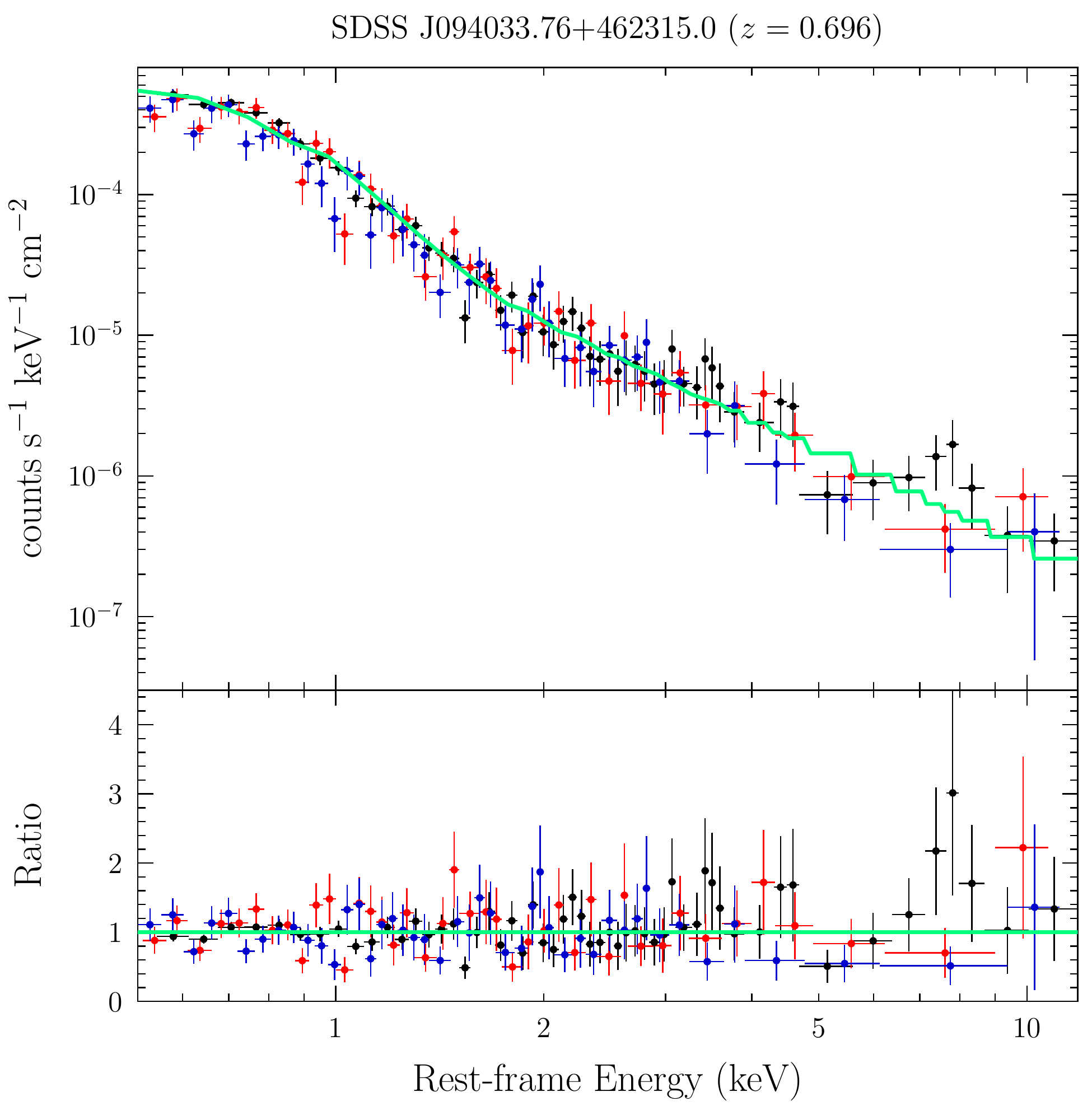}} \hfill\null

  \caption{(Continued)}
\end{figure*}

\clearpage

\begin{figure*}[!htbp]
  \ContinuedFloat
  \centering
  
  \subfigure{\includegraphics[width=0.4\textwidth]{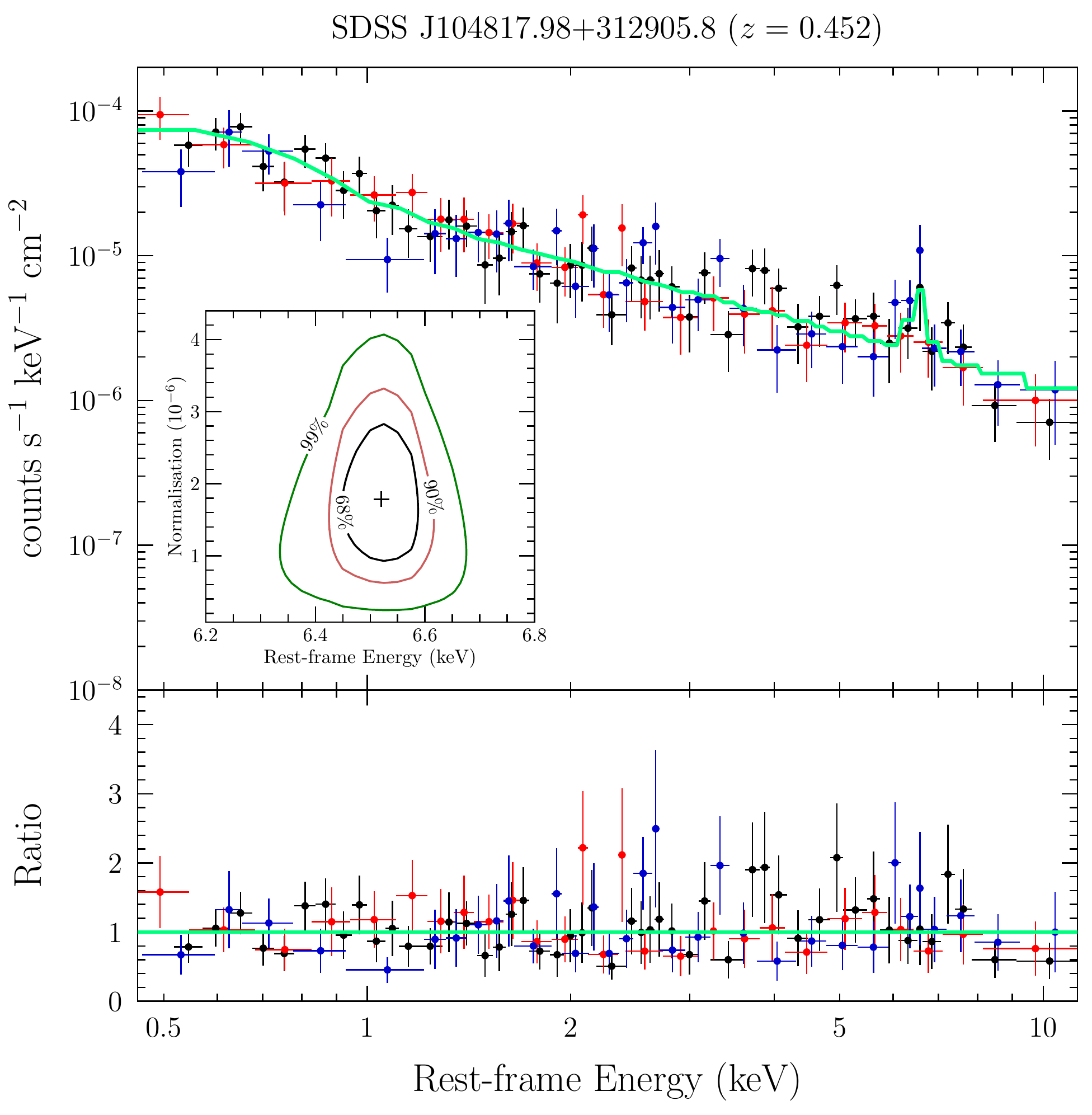}} \hfill
  \subfigure{\includegraphics[width=0.4\textwidth]{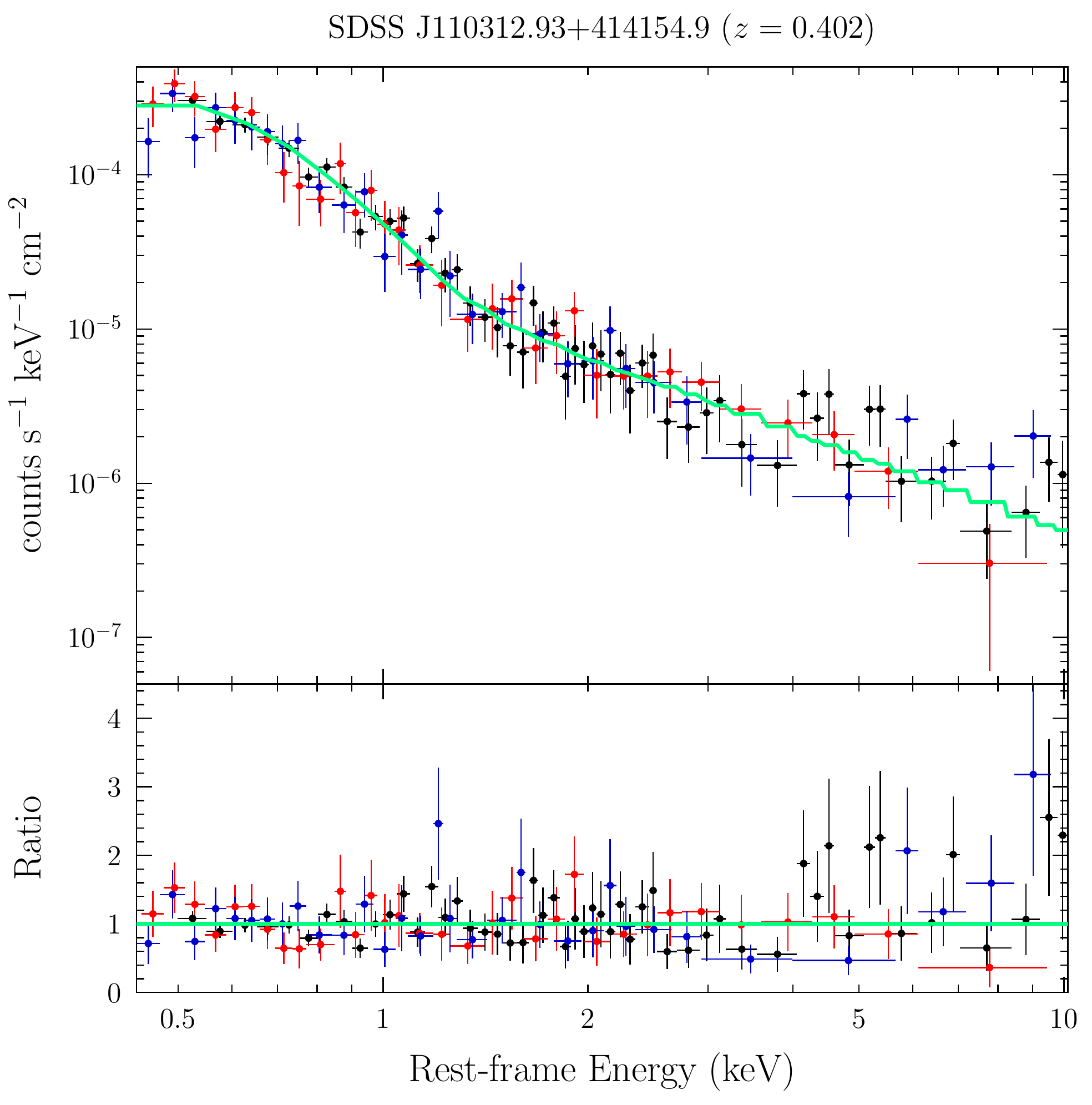}} \hfill\null
  
  \subfigure{\includegraphics[width=0.4\textwidth]{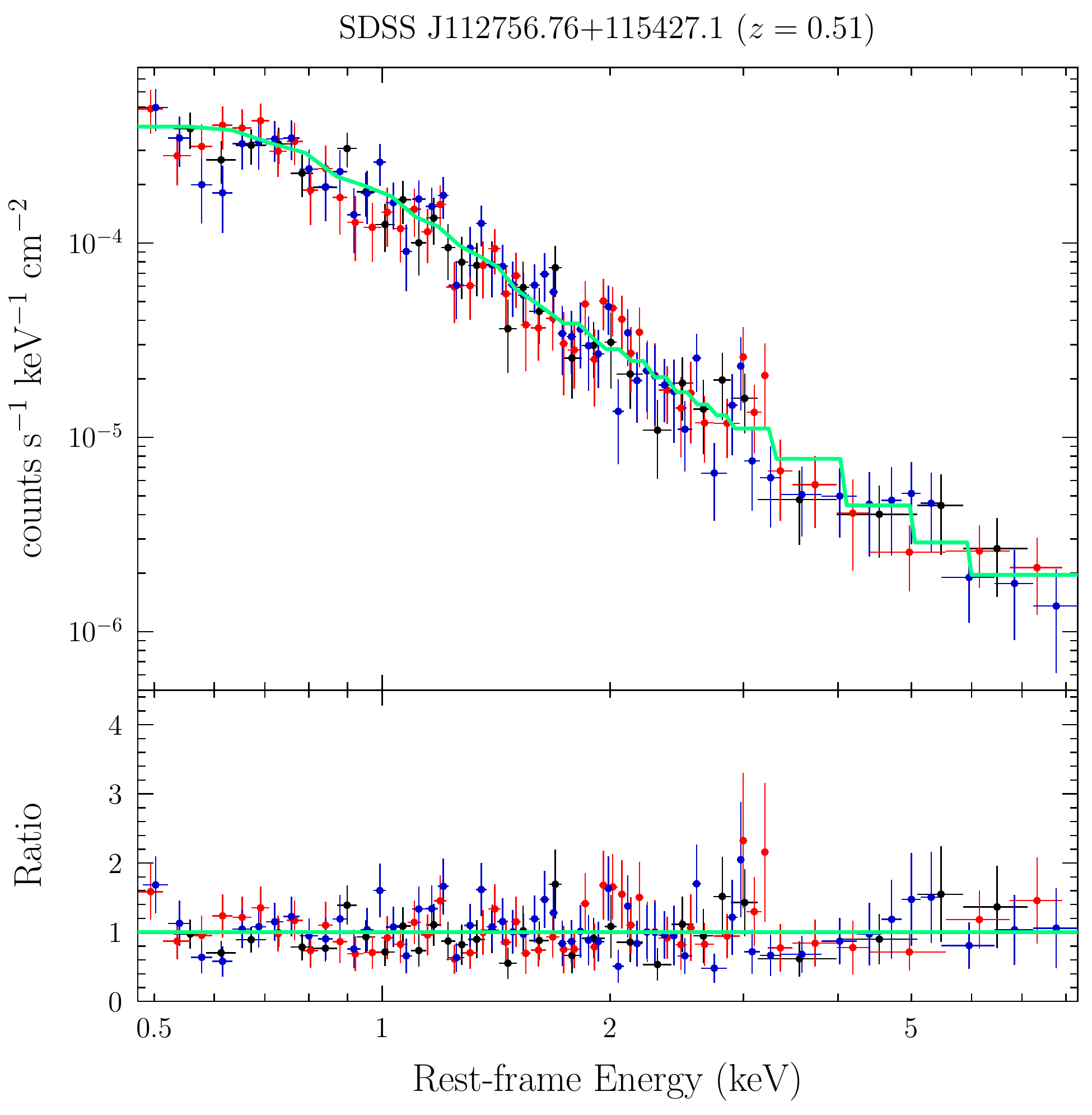}} \hfill
  \subfigure{\includegraphics[width=0.4\textwidth]{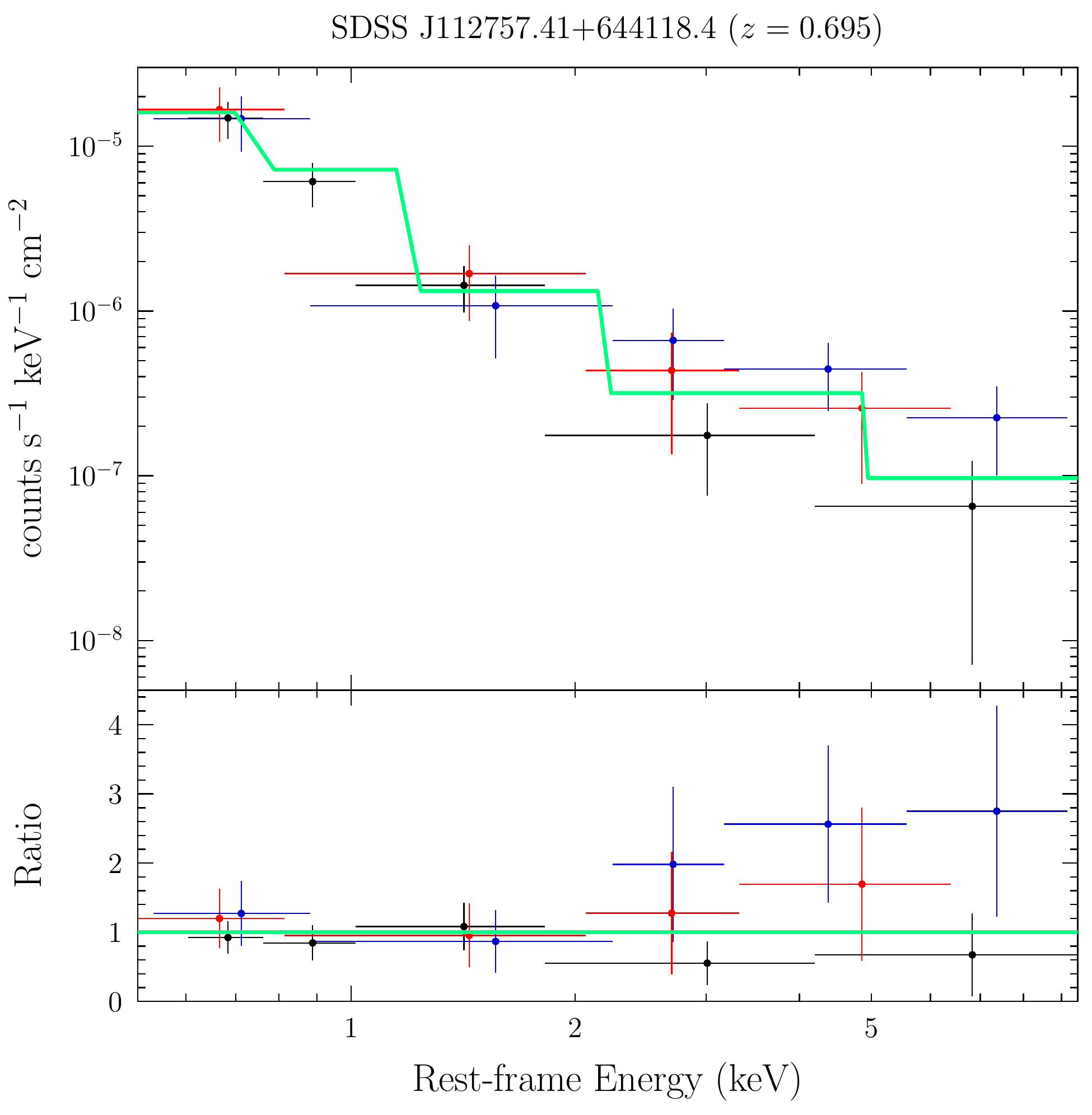}} \hfill\null
  
  \subfigure{\includegraphics[width=0.4\textwidth]{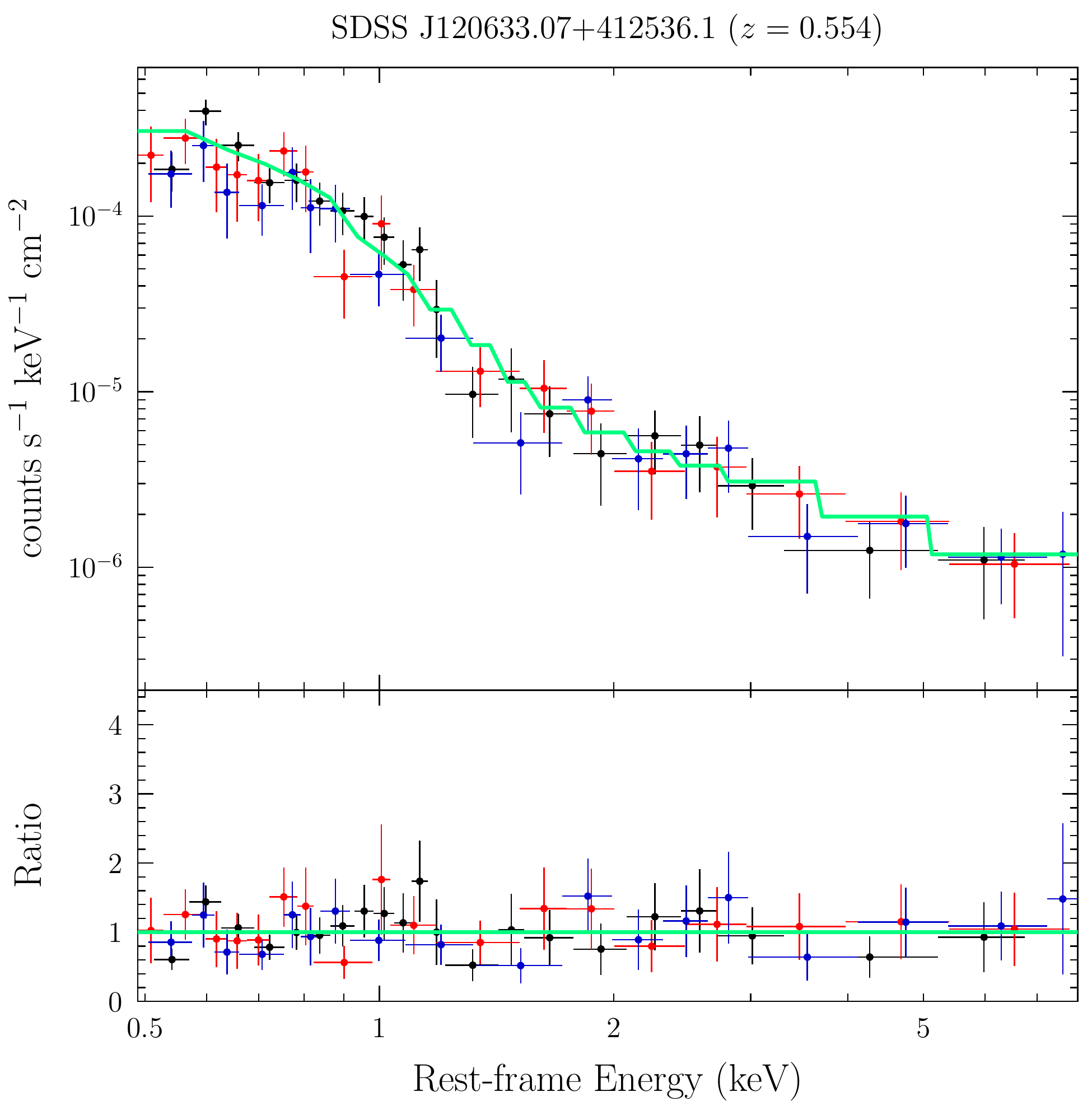}} \hfill
  \subfigure{\includegraphics[width=0.4\textwidth]{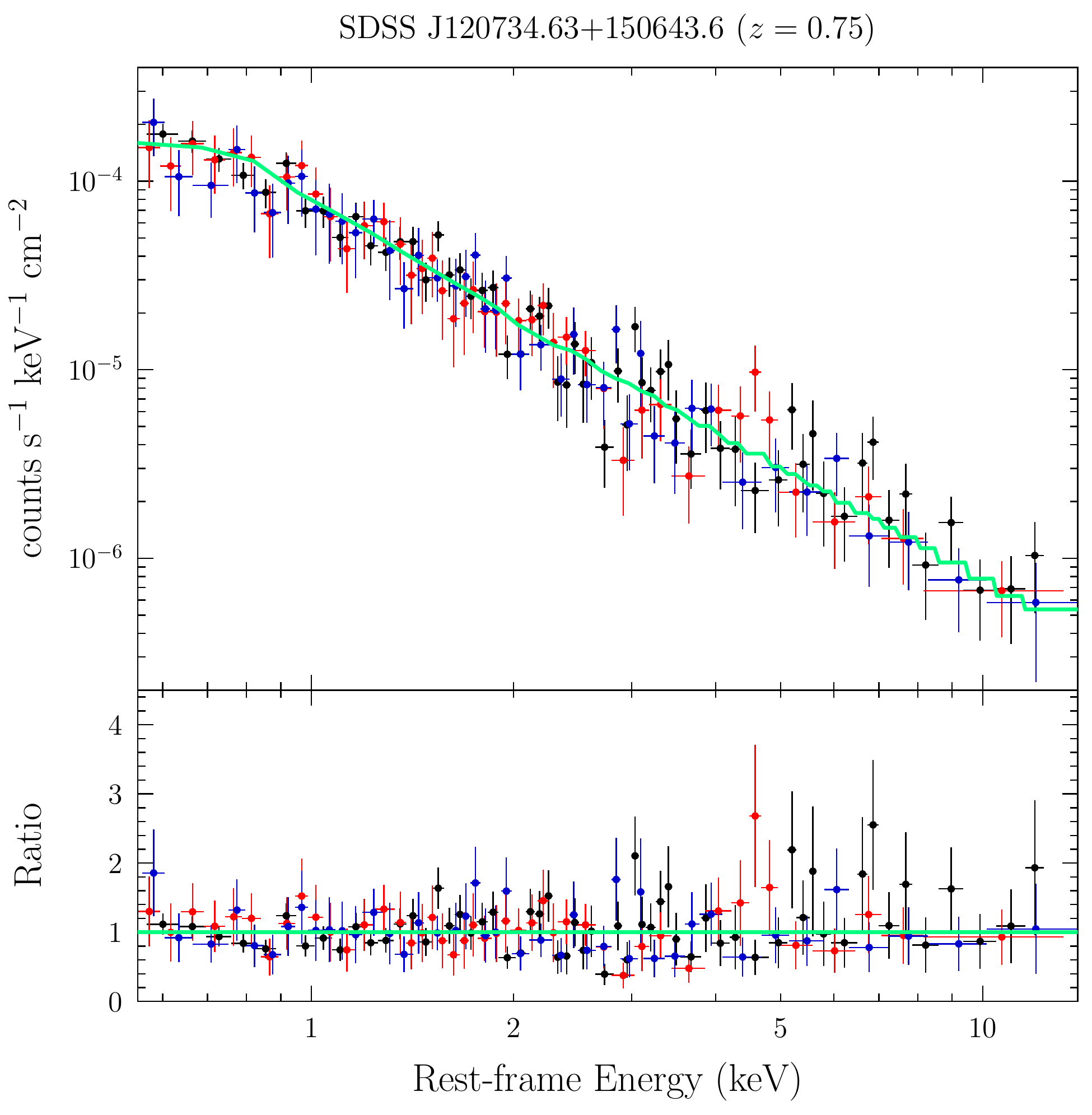}} \hfill\null
  
  \caption{(Continued)}
\end{figure*}

\begin{figure*}[!htbp]
  \ContinuedFloat
  \centering
  
  \subfigure{\includegraphics[width=0.4\textwidth]{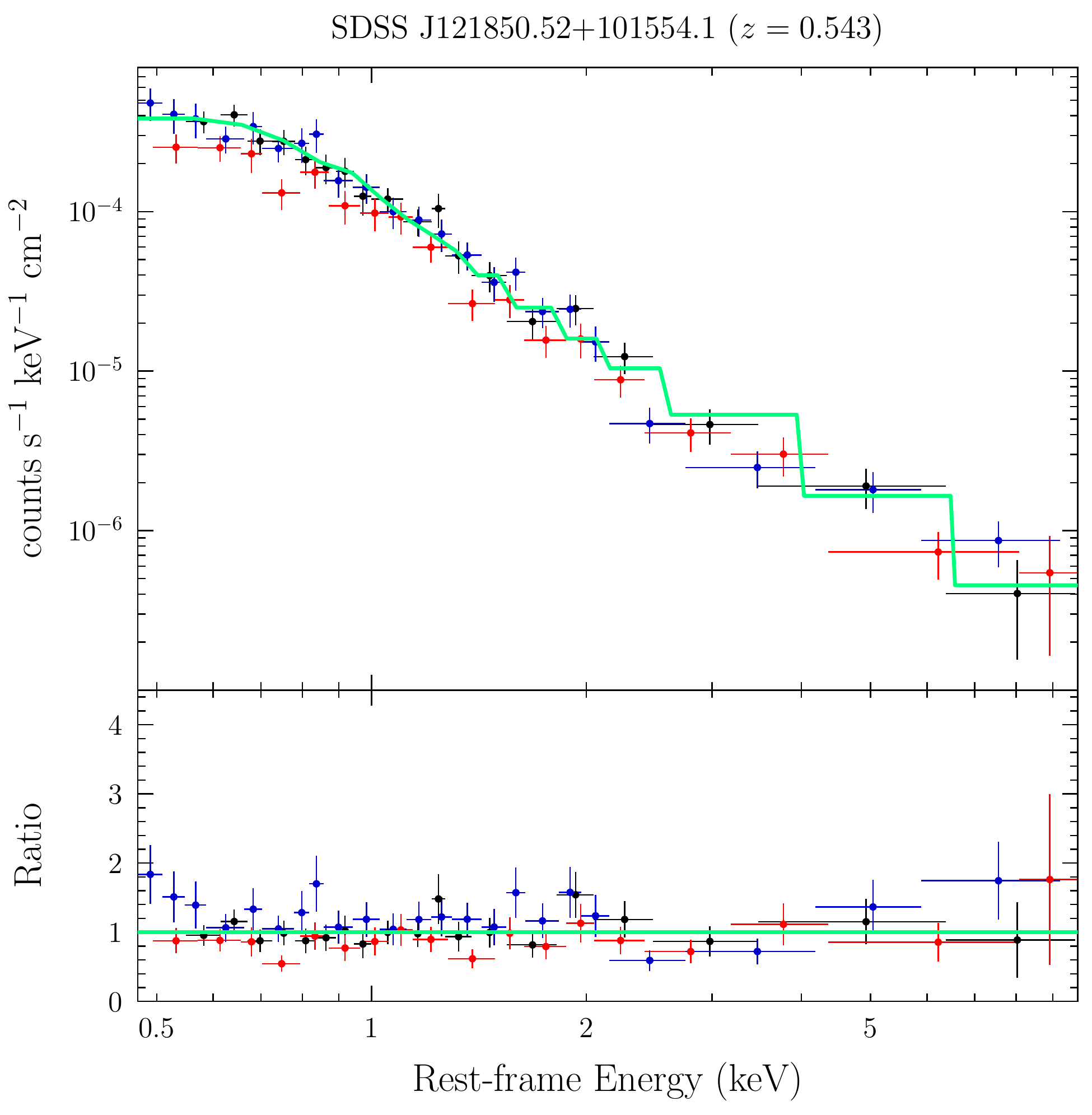}} \hfill
  \subfigure{\includegraphics[width=0.4\textwidth]{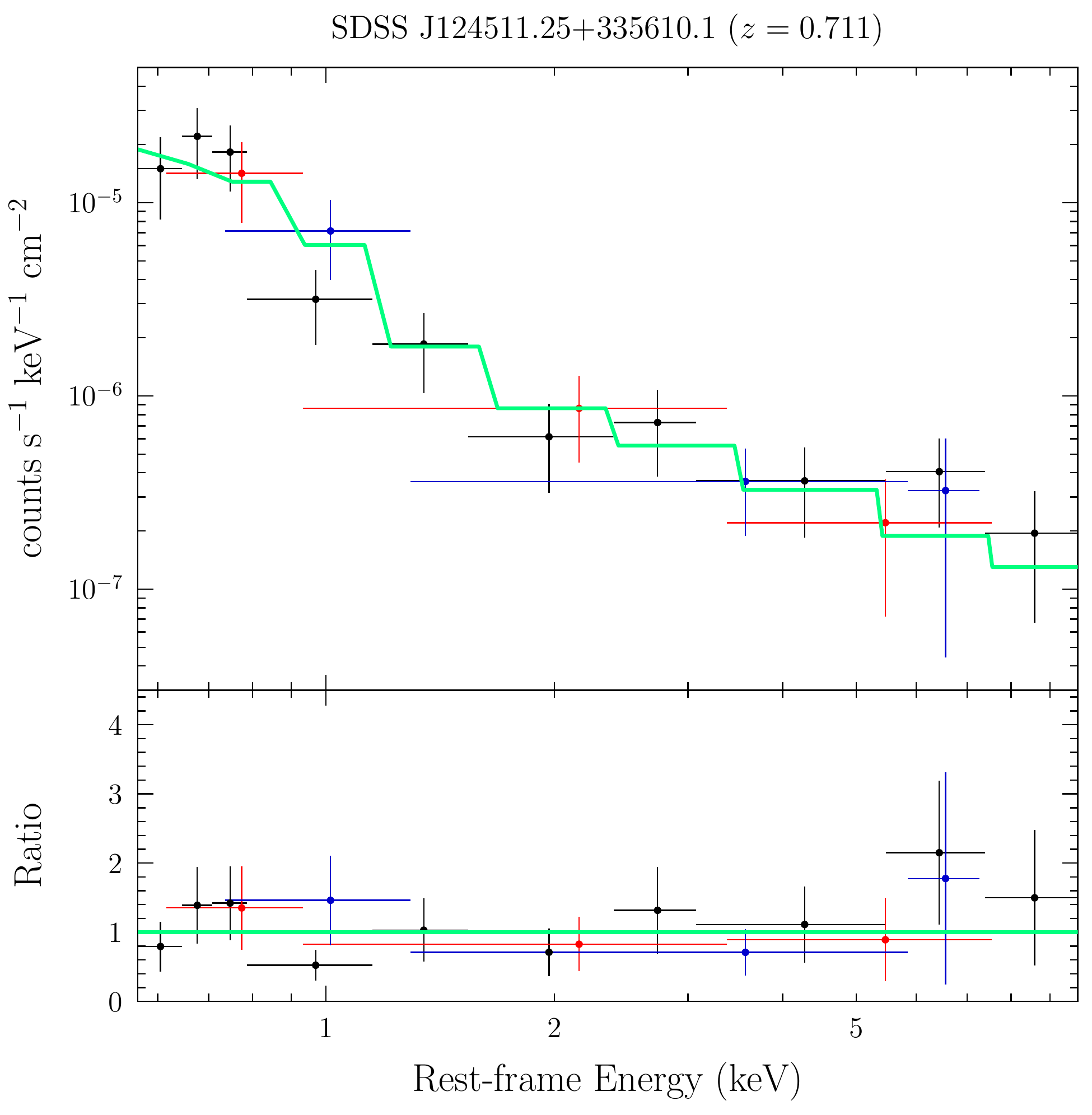}} \hfill\null
  
  \subfigure{\includegraphics[width=0.4\textwidth]{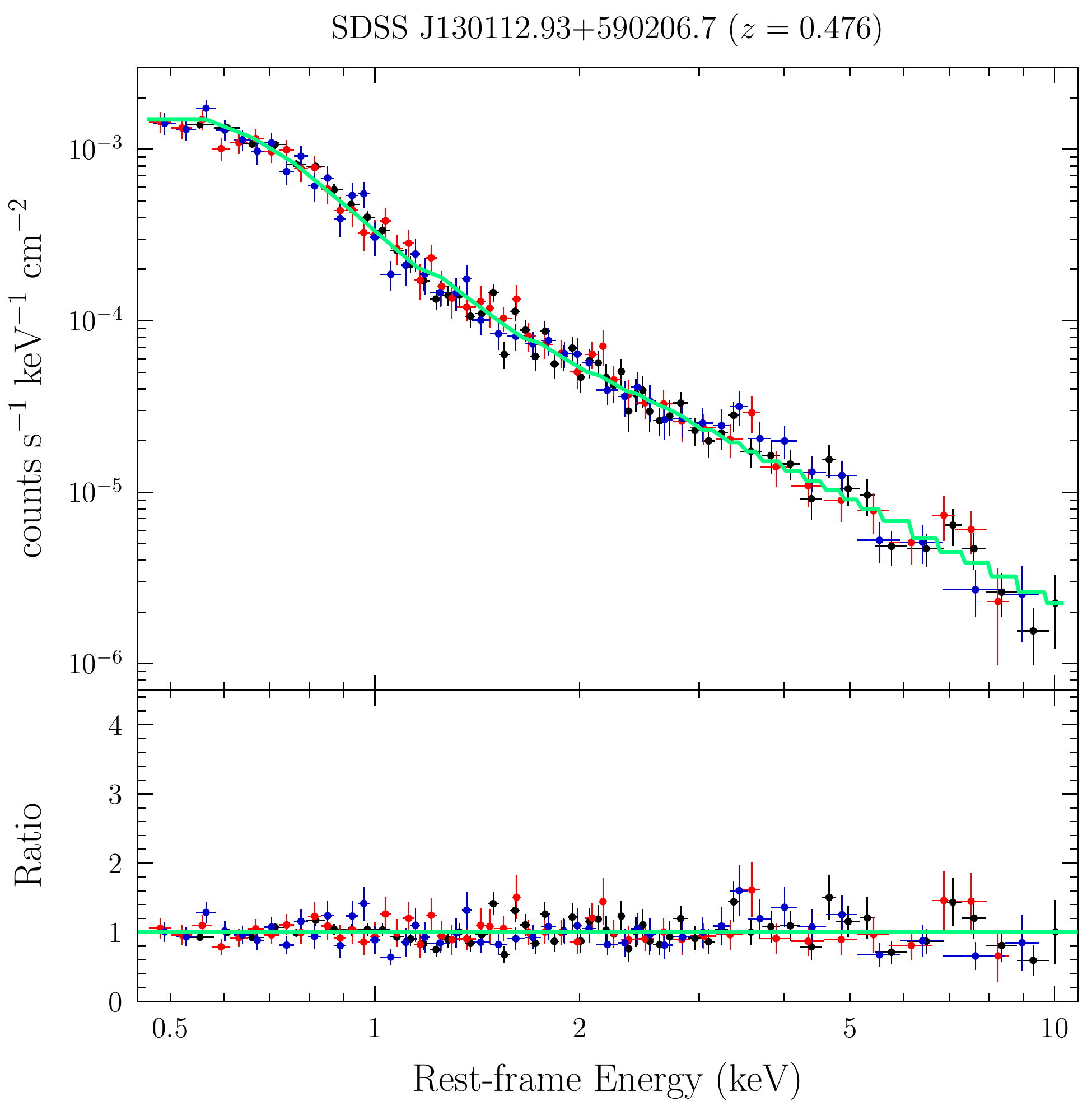}} \hfill
  \subfigure{\includegraphics[width=0.4\textwidth]{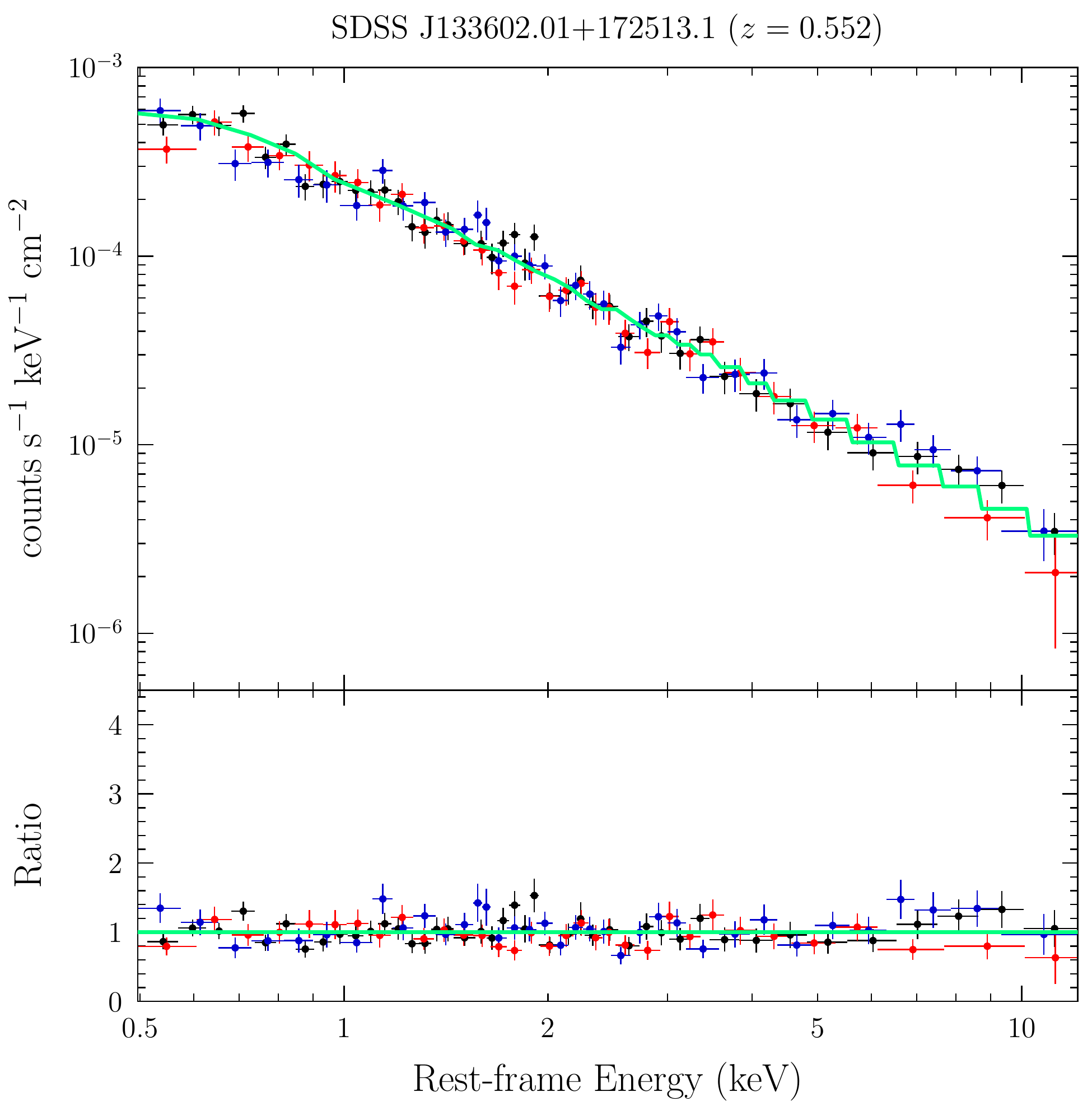}} \hfill\null
  
  \caption{\emph{XMM-Newton} spectra of the 14 high-\ledd\ quasars analysed in this paper.  For each source, the  top panel shows  the  X-ray spectrum  (data from pn, MOS1 and MOS2 are indicated as black, red and blue points, respectively) and the corresponding best-fit model (green solid line).  The bottom panel shows the ratios between the data points and the best-fit model, except for J0809+46, whose spectrum is dominated by an absorption feature in the soft X-rays. To highlight the presence of this component, for this source  the extension of the best-fitting hard X-ray continuum  power-law model to the soft X-ray portion is shown.
Furthermore, for the quasar J1048+31, the inset figure displays the confidence contour plot of normalisation against rest-frame energy of the Gaussian line component which accounts for  Fe K$\alpha$ emission.}
  \label{fig:spectra}
  
\end{figure*}

\section{Serendipitous sources}
\label{sec:serendipitous}

\begin{figure*}[ht]
  \centering
  \subfigure{\includegraphics[width=0.4\textwidth]{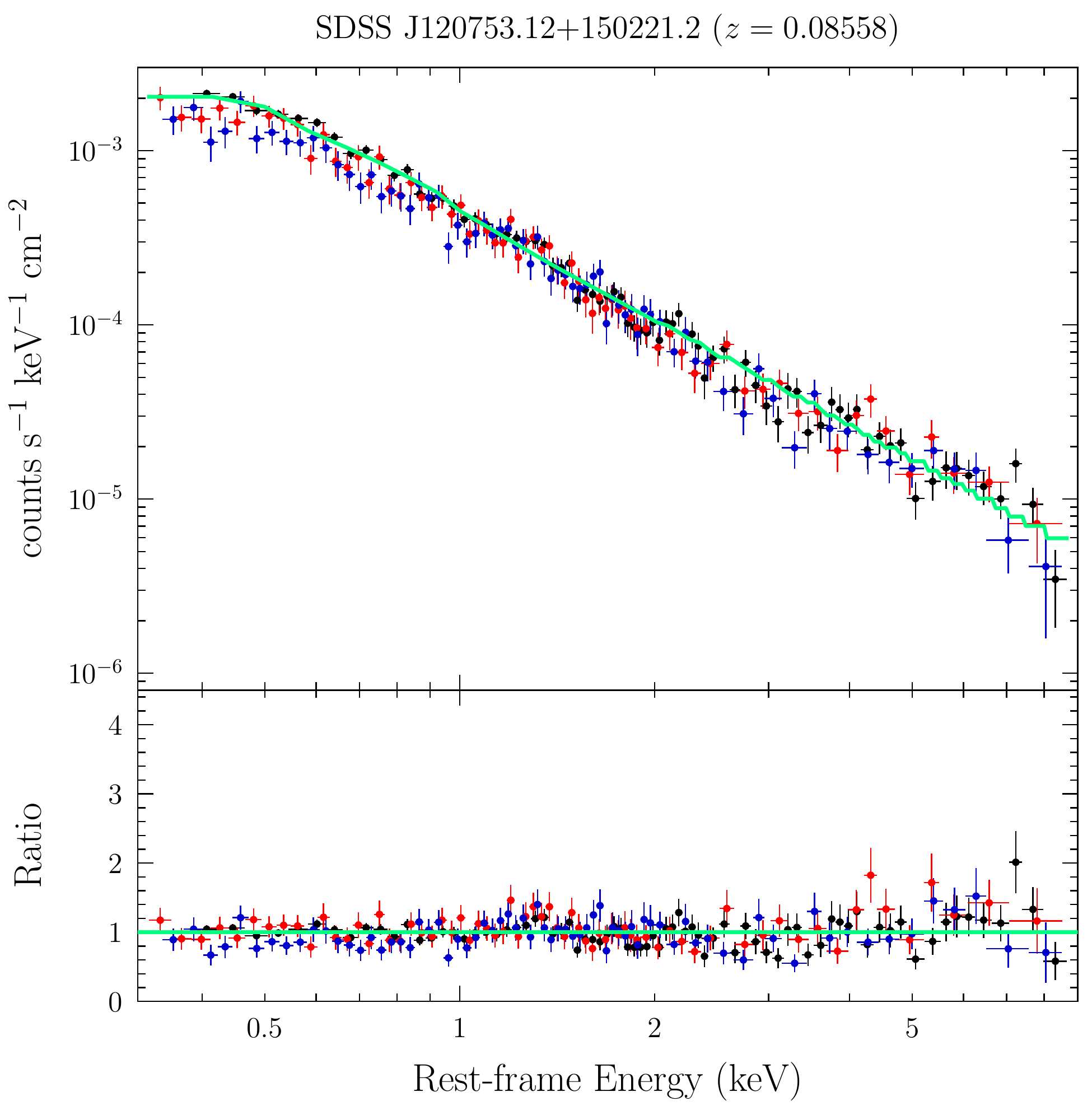}} \hfill
  \subfigure{\includegraphics[width=0.4\textwidth]{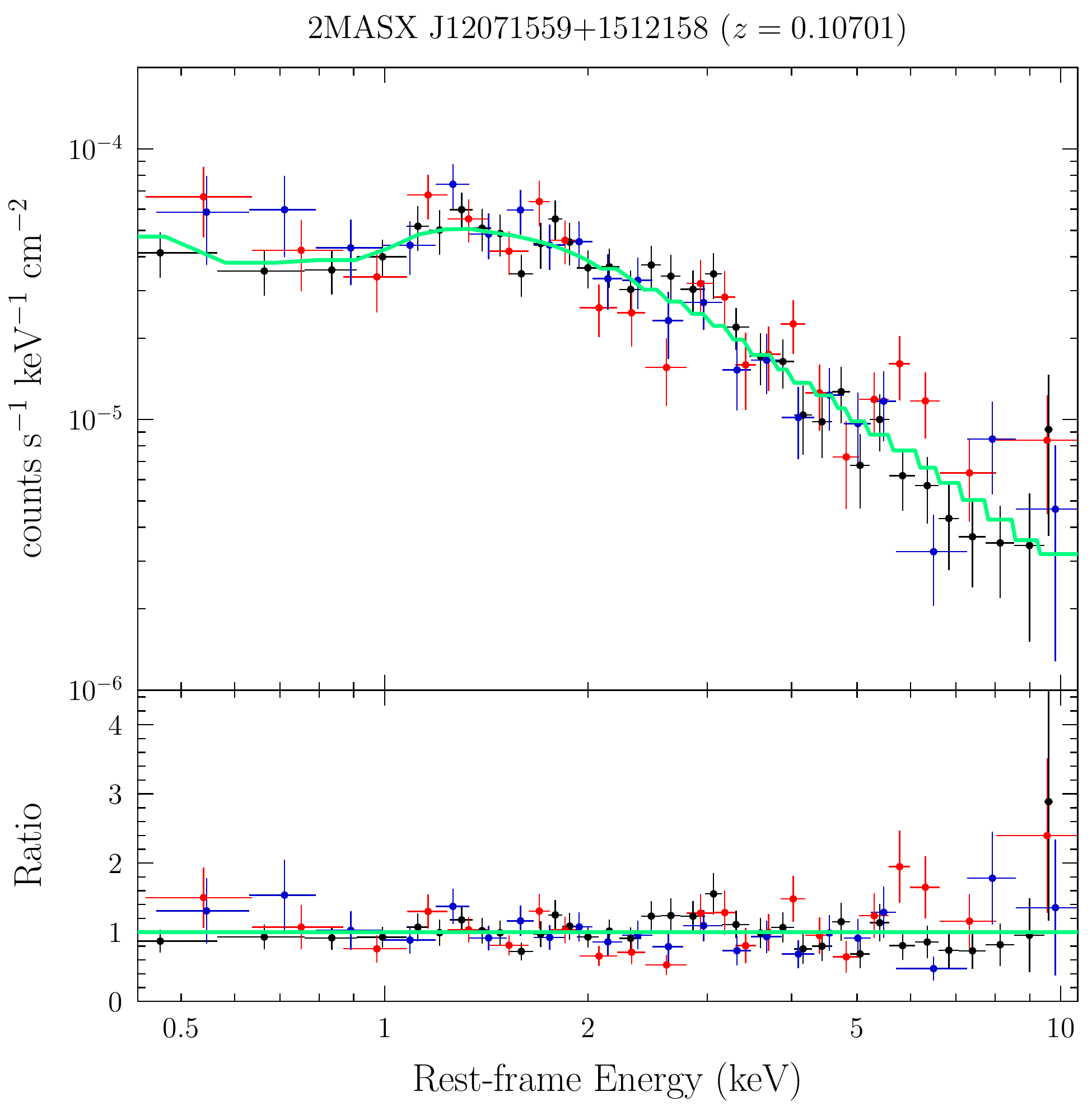}} \hfill\null
  
  \subfigure{\includegraphics[width=0.4\textwidth]{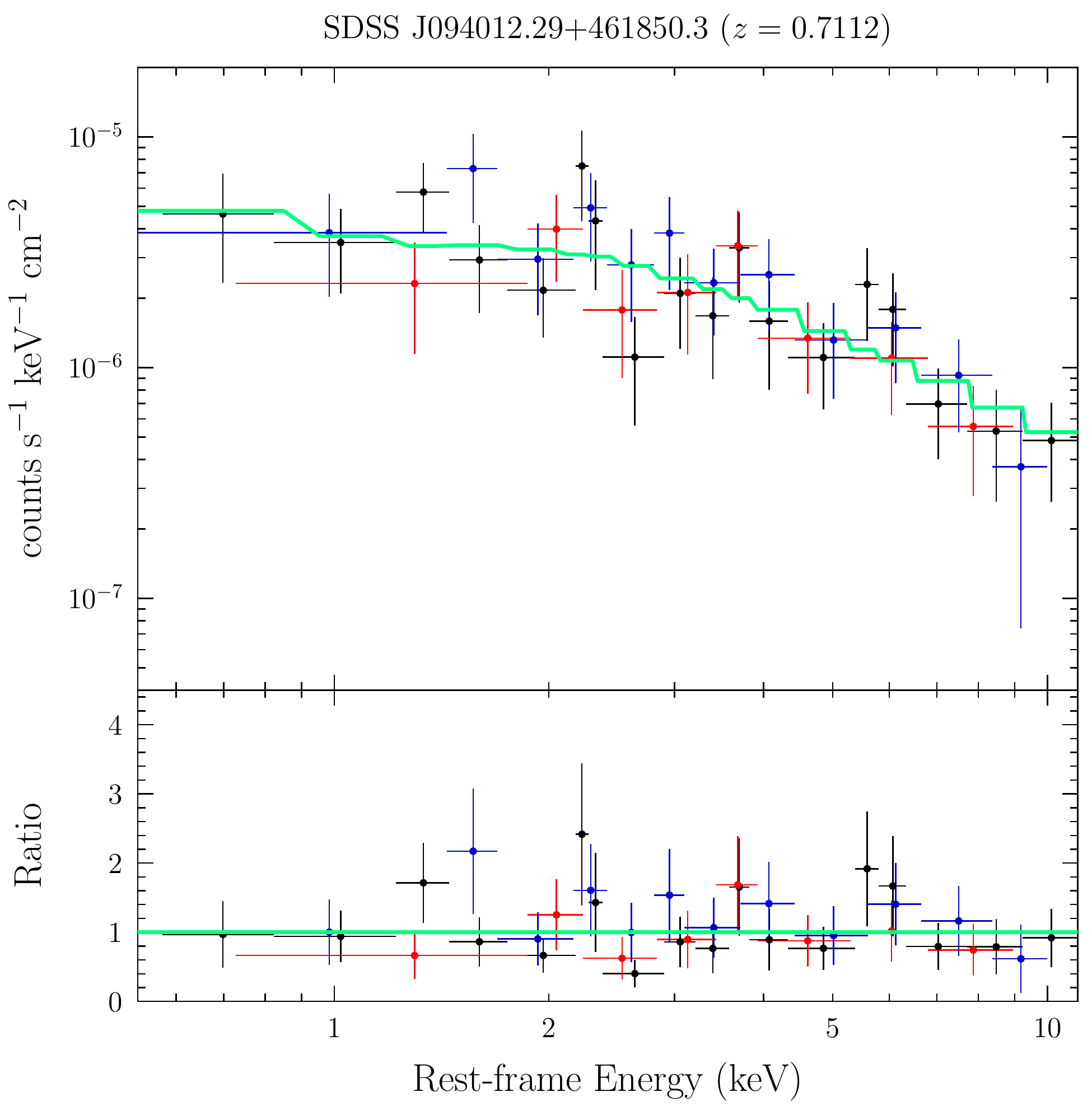}} \hfill
  \subfigure{\includegraphics[width=0.4\textwidth]{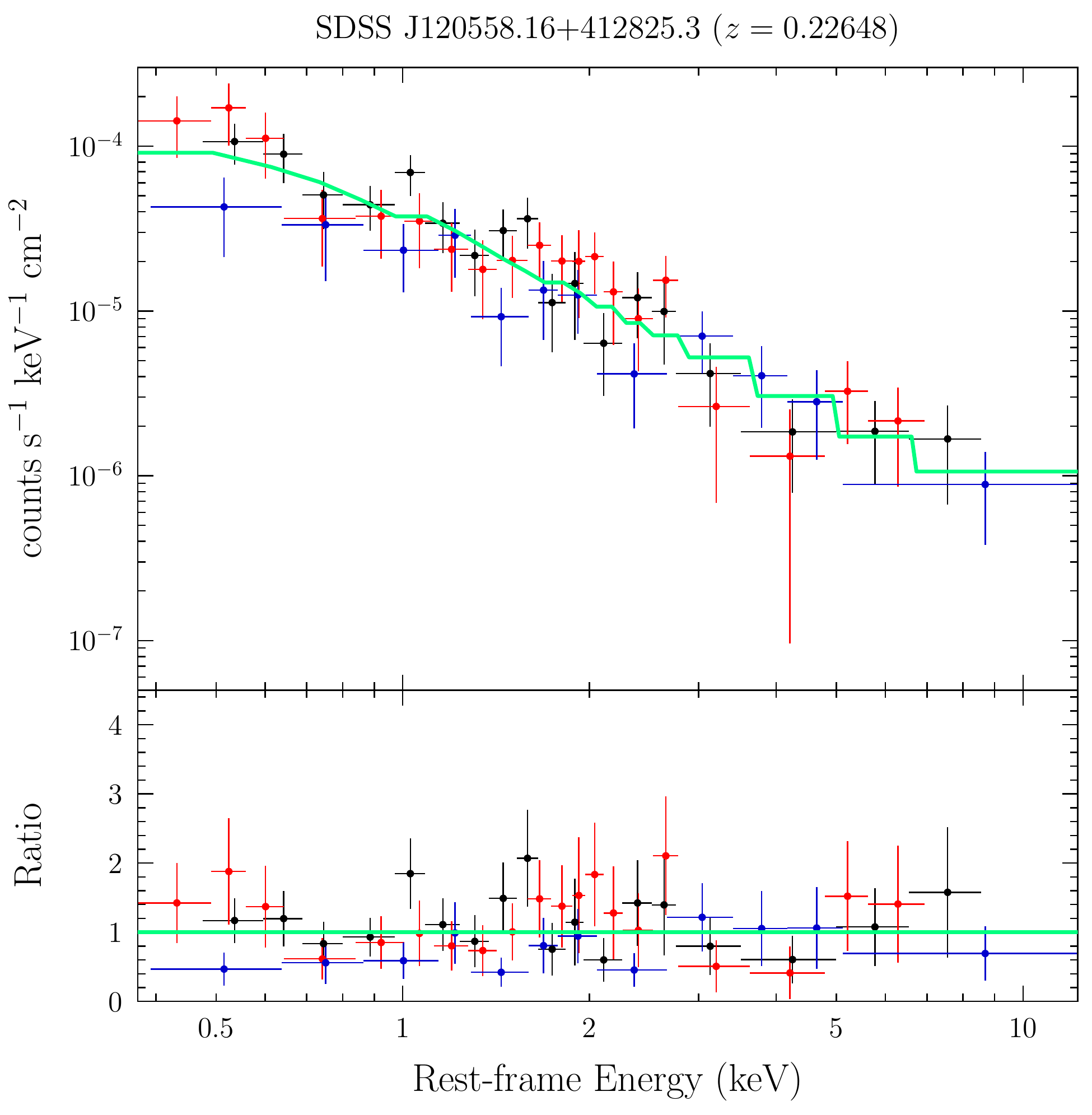}} \hfill\null
  \caption{\textit{XMM-Newton} spectra of the serendipitous sources described in Appendix \ref{sec:serendipitous}, with their corresponding best fit. Each top panel shows the X-ray spectrum and the best-fit line (in green). Each spectrum is rebinned for plotting purposes only. The bottom panels describe the ratio between the data points and the best-fit model. Data from EPIC-pn, MOS1 and MOS2 are shown in black, red and blue, respectively.}
  \label{fig:serendipitous}
\end{figure*}

In this section we briefly summarise the results of the X-ray spectral analysis of a small group of serendipitous X-ray sources with interesting X-ray properties and AGN optical classification available from the literature, located in the \emph{XMM-Newton} field of view of our targeted high-$\lambda_\mathrm{Edd}$ AGN.
Figure \ref{fig:serendipitous} shows their X-ray spectra along with the corresponding best fit model.

\textit{SDSS J120753.12+150221.2.} 
This source is an X-ray bright  Seyfert 1 galaxy  at redshift $z=0.08558$ (\citealt{oh2015}). Its X-ray spectrum is characterised by a soft excess with  $R_\mathrm{S/P}=0.16$. The broadband ($E=0.3-10$ keV observer frame) continuum is well described by a power law ($\Gamma=2.06\pm0.06$) modified by Galactic absorption, plus a blackbody component ($kT = 120_{-20}^{+10}$ eV).
In the $2-10$ keV energy interval, we measure a flux $F_\mathrm{2-10} \sim 10^{-12}$ erg cm$^{-2}$ s$^{-1}$ and a luminosity $L_\mathrm{2-10} \sim 1.9 \times 10^{43}$ erg s$^{-1}$. The net pn/MOS1/MOS2 photon counts in the broadband $E=0.3-10$ keV observer-frame energy interval are 6098/2060/1963.

\textit{2MASX J12071559+1512158.} 
This source is classified as a Seyfert 1 galaxy (\citealt{toba2014}) at redshift $z=0.10701$. Its soft X-ray spectrum is characterised by absorption below 1 keV rest frame. A good fit of the broadband continuum consists of a power law component with $\Gamma = 1.8\pm0.2$.
The absorption feature in the soft X-rays is modelled with a photoeletric absorber (\texttt{zwabs} component in XSPEC) with a column density of $N_\mathrm{H}= (0.7\pm0.2) \times 10^{22}$ cm$^{-2}$. 
The source has a flux $F_\mathrm{2-10} \sim 5.7 \times 10^{-13}$ erg cm$^{-2}$ s$^{-1}$ and a luminosity $L_\mathrm{2-10} \sim 1.6 \times 10^{43}$ erg s$^{-1}$.
The net pn/MOS1/MOS2 photon counts in the broadband $E=0.3-10$ keV observer-frame energy interval are 859/397/362.

\textit{SDSS J094012.29+461850.3.}
This QSO is located at redshift $z=0.7112$ (\citealt{paris2014}). To fit the X-ray broadband continuum, we first considered a simple model consisting of a power law modified by Galactic absorption. However, such a model resulted in a poor fit of the underlying continuum. We suspected the discrepancy could arise from the deficit below 2 keV rest frame. So we included a partial covering, cold absorption component in the model. 
The covering factor and column density measured with {\em XMM-Newton} are $C_\mathrm{f}=0.7_{-0.2}^{+0.1}$ and  $N_\mathrm{H} = 1.7_{-1.1}^{+1.5}\times10^{22}$ cm$^{-2}$, respectively. The primary continuum is described by a power law with photon index $\Gamma = 1.4\pm0.3$. 
We estimate a flux $F_\mathrm{2-10} \sim 7.1 \times 10^{-14}$ erg cm$^{-2}$ s$^{-1}$ and a luminosity $L_\mathrm{2-10} \sim 1.2 \times 10^{44}$ erg s$^{-1}$. 
The net pn/MOS1/MOS2 photon counts in the broadband $E=0.3-10$ keV observer-frame energy interval are 104/42/64.

\textit{SDSS J120558.16+412825.3.}
This source is a Seyfert 1 galaxy at redshift $z=0.22648$  (\citealt{toba2014}). The \emph{XMM-Newton} data do not require the presence of a soft excess and  a model consisting of a power law, modified by Galactic absorption, with photon index $\Gamma = 1.8 \pm 0.2$ provides a best fit to the X-ray spectrum.
We estimate a flux $F_\mathrm{2-10} \sim 1.2 \times 10^{-13}$ erg cm$^{-2}$ s$^{-1}$, which implies a luminosity $L_\mathrm{2-10} \sim 1.8 \times 10^{43}$ erg s$^{-1}$. The net pn/MOS1/MOS2 photon counts in the broadband $E=0.3-10$ keV observer-frame energy interval are 121/96/53.

\end{document}